\documentstyle[12pt,aps,epsf]{revtex}
\newcommand{\beq}{\begin{equation}}
\newcommand{\eeq}{\end{equation}}
\begin{document}

\title{INTRODUCTION TO CHAOS AND DIFFUSION}

\author{G.~Boffetta$^1$, G.~Lacorata$^2$ and A.~Vulpiani$^3$}
\address{$^1$Dipartimento di Fisica Generale and INFM,
Universit\`a di Torino, \\
Via Pietro Giuria 1, 10125 Torino, Italy}
\address{$^2$ CNR - Istituto di Scienze dell'Atmosfera e del Clima, 
Sezione di Lecce, \\
Str. Pr. Lecce-Monteroni, 73100 Lecce, Italy}
\address{$^3$ Dipartimento di Fisica, INFM, UdR and CSM, 
Universit\`a di  Roma "la Sapienza", \\
Piazzale Aldo Moro 5, 00185 Roma, Italy}

\maketitle

%%%%%%%%%%%%%%%%%%%%%%%%%%%%%%%%%%%%%%%%%%%%%%%%%%%%%%%%%%%%%%
\section{Introduction}
\label{sec:1}

This contribution is relative to the opening lectures of the 
ISSAOS 2001 summer school and it  has the aim to provide the reader 
with some concepts and techniques concerning chaotic dynamics
and transport processes in fluids.
Our intention is twofold: to give a self-consistent introduction
to chaos and diffusion, and to offer a guide for the reading of 
the rest of this volume.

In the following Section we present some basic elements
of the chaotic dynamical systems theory, as the Lyapunov
exponents and the Kolmogorov-Sinai entropy. 
The third Section is devoted to Lagrangian chaos in fluids.
The last Section contains an introduction to  diffusion and
transport processes, with particular emphasis
on the treatment of non-ideal cases.

%%%%%%%%%%%%%%%%%%%%%%%%%%%%%%%%%%%%%%%%%%%%%%%%%%%%%%%%%%%%%%
\section{Some basic elements of dynamical systems}
\label{sec:2}

\begin{figure}[hbt]
\epsfxsize=12truecm
\centerline{\epsfbox{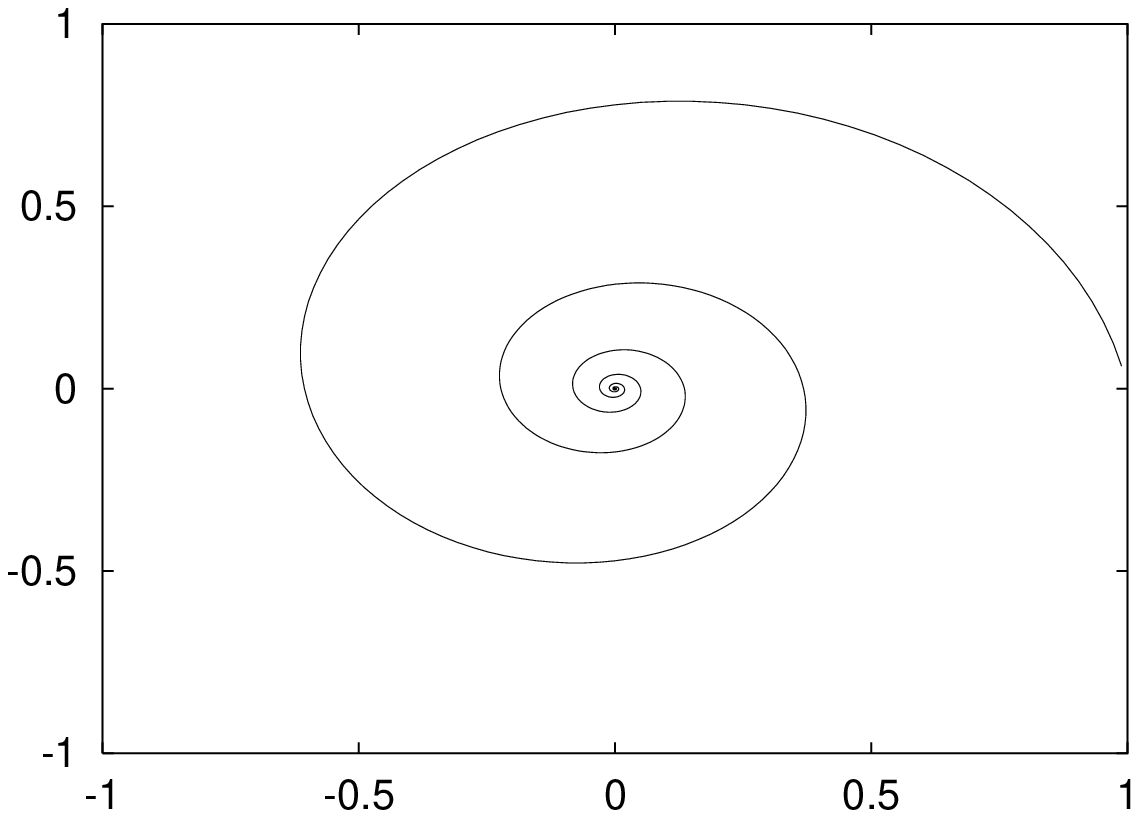}}
\epsfxsize=12truecm
\centerline{\epsfbox{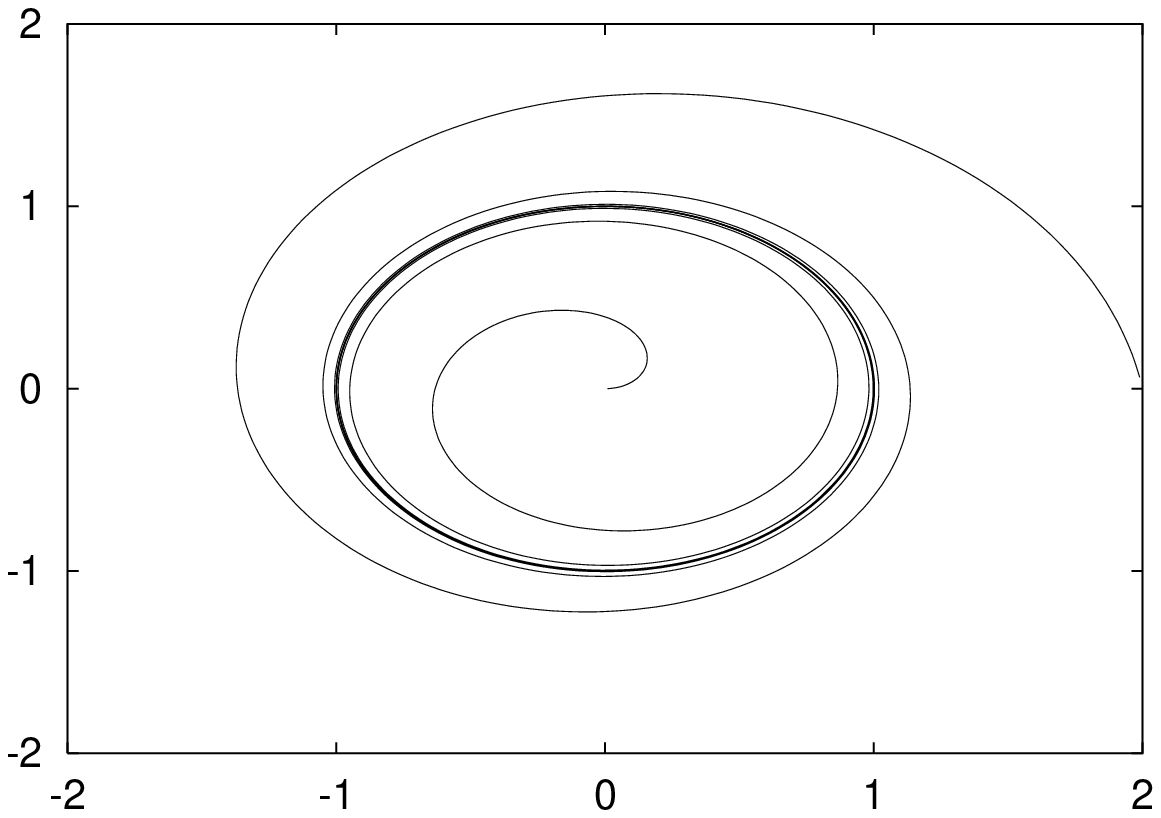}}
\caption{Examples of regular attractors: fixed point (above) and 
limit cycle (below).
}
\label{fig:regular}
\end{figure}

A dynamical system may be defined as a deterministic rule for the time 
evolution of state observables. Well known examples are the ordinary 
differential equations (ODE) in which time is continuous:
\begin{equation}
{d {\bf x}(t) \over dt} = {\bf f}({\bf x}(t)),  \;\;\;\;\; 
{\bf x},{\bf f} \in {\rm I\!R}^d;
\label{eq:ode}
\end{equation}
and maps in which time is discrete:
\begin{equation}
{\bf x}(t+1)={\bf g}({\bf x}(t)), \;\;\;\;\;
{\bf x},{\bf g} \in {\rm I\!R}^d. 
\label{eq:maps}
\end{equation}
In the case of maps, the evolution law is straightforward: from 
${\bf x}(0)$ one computes ${\bf x}(1)$, and then ${\bf x}(2)$ and 
so on. For ODE's, under rather general assumptions on {\bf f}, from 
an initial condition ${\bf x}(0)$ one has a unique trajectory ${\bf x}(t)$ 
for $t>0$ \cite{Ott93}. Examples of regular behaviors (e.g. stable fixed points, 
limit cycles) are well known, see Figure~\ref{fig:regular}. 

A rather natural question is the possible existence of less regular 
behaviors i.e. different from stable fixed points, periodic or 
quasi-periodic motion. 

\begin{figure}[hbt]
\epsfxsize=12truecm
\centerline{\epsfbox{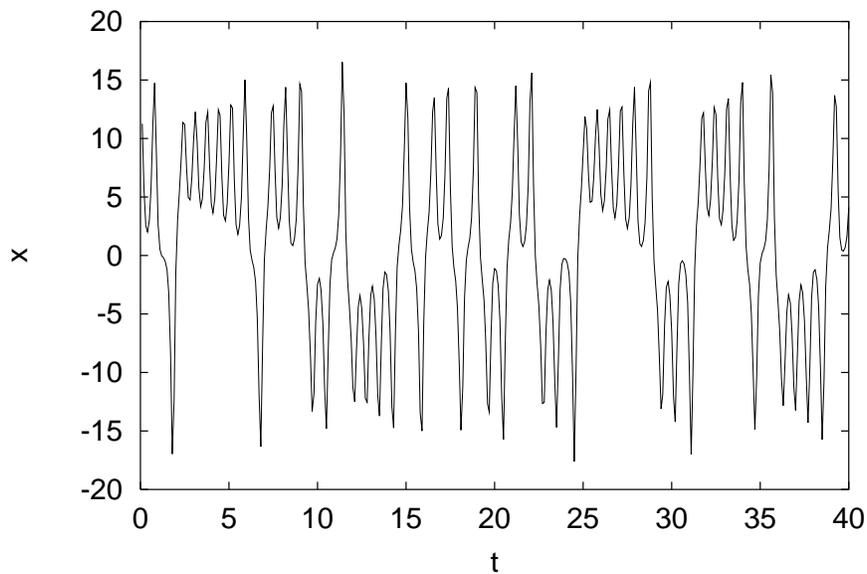}}
\caption{Example of aperiodic signal: the $x$ variable of the 
Lorenz system (\ref{eq:lorenz}) as function of time $t$, for $r = 28$.}
\label{fig:aperiod}
\end{figure}

After the seminal works of Poincar\'e, Lorenz and H\'enon 
(to cite only the most eminent ones) it is now well established that the 
so called chaotic behavior is ubiquitous. As a relevant system, originated 
in the geophysical context, we mention the celebrated Lorenz model 
\cite{L63}:
%\begin{equation}
%\left\{
\begin{eqnarray}
{dx \over dt} & = & -\sigma (x - y) \nonumber\\
{dy \over dt} & = & -xz+rx-y \label{eq:lorenz}\\
{dz \over dt} & = & xy-bz \nonumber
\end{eqnarray}
%\right.
%\label{eq:lorenz}
%\end{equation}
This system is related to the Rayleigh-Benard convection under very 
crude approximations. The quantity $x$ is proportional the circulatory 
fluid particle velocity; the quantities $y$ and $z$ are related to 
the temperature profile; $\sigma$, $b$ and $r$ are dimensionless 
parameters. Lorenz studied the case with $\sigma=10$ and $b=8/3$ at varying 
$r$ (which is proportional to the Rayleigh number). 
It is easy to see by linear analysis that the fixed point $(0,0,0)$ is 
stable for $r < 1$. For $r > 1$ it becomes unstable and two new 
fixed points appear:
\begin{equation}
C_{+,-}=(\pm \sqrt{b(r-1)}, \pm \sqrt{b(r-1)}, r-1), 
\label{eq:fixedpts}
\end{equation}
these are stable for $r < r_c = 24.74$.  
A nontrivial behavior, i.e. non periodic, is present for $r > r_c$, 
as is shown in Figure~\ref{fig:aperiod}.

\begin{figure}[hbt]
\epsfxsize=12truecm
\centerline{\epsfbox{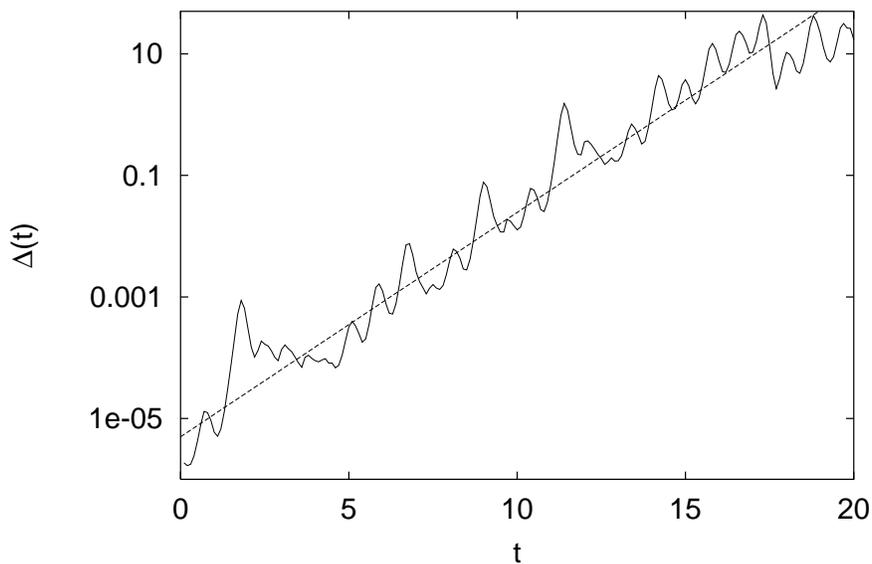}}
\caption{The growth of a generic infinitesimal 
perturbation $\Delta(t)$ in the Lorenz model (\ref{eq:lorenz}) as 
function of time $t$, for $r=28$. The dashed line represent the 
exponential growth $\Delta(0) \exp{\lambda_1 t}$.
}
\label{fig:pert}
\end{figure}

In this ``strange'', chaotic regime one has the so called sensitive 
dependence on initial conditions. Consider two trajectories, 
${\bf x}(t)$ and ${\bf x}'(t)$, initially very close and denote with 
$\Delta(t)=||{\bf x}'(t)-{\bf x}(t)||$ their separation.
Chaotic behavior means that if
$\Delta(0) \to 0$, then as $t \to \infty$ one has
$\Delta(t) \sim \Delta(0) \exp{\lambda_1 t}$, with $\lambda_1 > 0$, 
see Figure \ref{fig:pert}.

Let us notice that, because of its chaotic behavior and its dissipative 
nature, i.e.
\begin{equation}
{\partial \dot x \over \partial x} +
{\partial \dot y \over \partial y} + 
{\partial \dot z \over \partial z} < 0,
\label{eq:divlorenz}
\end{equation} 
the attractor of the Lorenz system cannot be a smooth surface. 
Indeed the attractor has a self-similar structure with a fractal dimension 
between $2$ and $3$. 
The Lorenz model (which had an important historical relevance 
in the development of chaos theory) is now considered a paradigmatic 
example of a chaotic system.

%%%%%%%%%%%%%%%%%%%%%%%%%%%%%%%%%%%%%%%%%%%%%%%%%%%%%
\subsection{Lyapunov exponents}
\label{sec:2.1}

The sensitive dependence on the initial conditions can be formalized 
in order to give it a quantitative characterization. 
The main growth rate of trajectory separation is measured by
the first (or maximum) Lyapunov exponent, defined as
\begin{equation}
\lambda_1 = \lim_{t \to \infty} \lim_{\Delta(0) \to 0} 
{1 \over t} \ln {\Delta(t) \over \Delta(0)}, 
\label{eq:lyap1}
\end{equation} 
As long as $\Delta(t)$ remains sufficiently small (i.e. infinitesimal, 
strictly speaking), one can regard the separation as a tangent vector 
${\bf z}(t)$ whose time evolution is
\begin{equation}
{dz_i \over dt} = \sum_{j=1}^{d} 
{\partial f_i \over \partial x_j}|_{{\bf x}(t)} \cdot z_j,
\label{eq:tangent}
\end{equation}
and, therefore,
\begin{equation}
\lambda_1 = \lim_{t \to \infty} 
{1 \over t} \ln {||{\bf z}(t)|| \over ||{\bf z}(0)||}.
\label{eq:lambdatang}
\end{equation}
In principle, $\lambda_1$ may depend on the initial condition ${\bf x}(0)$,
but this dependence disappears for ergodic systems. In general there exist 
as many Lyapunov exponents, conventionally written in decreasing order  
$\lambda_1 \ge \lambda_2  \ge \lambda_3 \ge ...$, as 
the independent coordinates of the phase space \cite{BGGS80}.
Without entering the details, one can define the sum of the first
$k$ Lyapunov exponents as the growth rate of an infinitesimal 
$k-$dimensional volume in the phase space.
In particular, $\lambda_1$ is the growth rate of material lines, 
$\lambda_1+\lambda_2$ is the growth rate of $2D$ surfaces, and so on.  
A numerical widely used efficient method is due to Benettin et al. 
\cite{BGGS80}. 

It must be observed that, after a transient, 
the growth rate of any generic small perturbation (i.e. distance between two 
initially close trajectories) is measured by the 
first (maximum) Lyapunov exponent $\lambda_1$, and $\lambda_1 > 0$ means 
chaos. In such a case, the state of the system is unpredictable on long 
times. Indeed, if we want to predict the state with a certain tolerance 
$\Delta$ then our forecast cannot be pushed over a certain time interval 
 $T_P$, called predictability time, given by:
\begin{equation}
T_P \sim {1 \over \lambda_1} \ln {\Delta \over \Delta(0)}. \label{eq:predtime}
\end{equation}
The above relation shows that $T_P$ is basically determined by $1/\lambda_1$, 
seen its weak dependence on the ratio $\Delta/\Delta(0)$. 
To be precise one must state that, for a series of reasons,
relation (\ref{eq:predtime}) is too simple to be of actual relevance 
\cite{boffi02}. 

%%%%%%%%%%%%%%%%%%%%%%%%%%%%%%%%%%%%%%%%%%%%%%%%%%%%%%%%%%%%%%%%%%%%
\subsection{The Kolmogorov-S entropy}
\label{sec:2.2}

Deterministic chaotic systems, because of their irregular behavior, 
have many aspects in common with stochastic processes. 
The idea of using stochastic processes to mimic chaotic behavior, therefore, 
is rather natural \cite{chirikov79,B84}. One of the most relevant 
and successful approaches is symbolic dynamics \cite{BS93}.
For the sake of simplicity let us consider a discrete time dynamical system. 
One can introduce a partition ${\cal A}$ of the phase space formed by 
N disjoint sets $A_1, ..., A_N$. From any initial condition one has a 
trajectory
\begin{equation}
{\bf x}(0) \to {\bf x}(1), {\bf x}(2), ..., {\bf x}(n), ...
\label{eq:traj}
\end{equation}
dependently on the partition element visited, the trajectory (\ref{eq:traj}), 
is associated to a symbolic sequence
\begin{equation}
{\bf x}(0) \to i_1, i_2, ..., i_n, ...
\label{eq:symbol}
\end{equation}
where $i_n$ ($i_n \in (1,2,...,N)$) means that 
${\bf x}(n) \in A_{i_n}$ at the step $n$, for $n=1,2,...$.
The coarse-grained properties of chaotic trajectories are therefore studied 
through the discrete time process (\ref{eq:symbol}). 

An important 
characterization of symbolic dynamics is given by the Kolmogorov-Sinai 
 (K-S) entropy, defined as follows. Let $C_n = (i_1, i_2, ..., i_n)$ be a 
generic ``word'' of size $n$ and $P(C_n)$ its occurrence probability,
the quantity
\begin{equation}
H_n = \sup_{A} [-\sum_{C_n} P(C_n) \ln P(C_n)] 
\label{eq:block}
\end{equation}
is called block entropy of the $n$-sequences, and it is computed 
by taking the largest value over all possible partitions. In the 
limit of infinitely long sequences, the asymptotic entropy increment 
\begin{equation}
h_{KS} = \lim_{n \to \infty} H_{n+1} - H_n
\label{eq:KS}
\end{equation}
is called Kolmogorov-Sinai entropy. The difference $H_{n+1} - H_n$ 
has the intuitive meaning of  
average information gain supplied by the $(n+1)-$th symbol, provided that 
the previous $n$ symbols are known. K-S entropy has an important
connection with the positive Lyapunov exponents of the system \cite{Ott93}:
\begin{equation}
h_{KS} = \sum_{\lambda_i > 0} \lambda_i
\label{eq:pesin}
\end{equation}
In particular, for low-dimensional chaotic systems for which only one 
Lyapunov exponent is positive, one has $h_{KS}=\lambda_1$. 

We observe that in (\ref{eq:block}) there is a technical difficulty,
i.e. taking the sup over all the possible partitions. However,
sometimes there exits a special partition, called generating partition, 
for which 
one finds that $H_n$ coincides with its superior bound. 
Unfortunately the generating partition is often hard to 
find, even admitting that it exist. 
Nevertheless, given a certain partition, chosen by 
physical intuition, the statistical properties of the related symbol 
sequences can give information on the dynamical system beneath. 
For example, if the probability of observing a symbol (state) depends only 
by the knowledge of the immediately preceding symbol, the symbolic process 
is called a Markov chain and all the statistical properties are determined 
by the transition matrix elements $W_{ij}$ giving the probability of 
observing a transition $i \to j$ in one time step. If the memory of 
the system extends far beyond the time step between two consecutive symbols, 
and the occurrence probability of a symbol depends on $k$ preceding steps, 
the process is called Markov process of order $k$ and, in principle, 
a $k$ rank tensor would be required to describe the dynamical system with 
 good accuracy. It is possible to demonstrate that if $H_{n+1}-H_n=h_{KS}$ 
for $n \ge k+1$, $k$ is the (minimum) order of the required Markov process 
 \cite{K57}. It has to be pointed out, however, that to know  
the order of the suitable Markov process we need is of no practical utility 
if $k \gg 1$. 

For applications of the Markovian approach to geophysical systems see \cite{CLVZ99} 
 and the contributions by Abel et al. and by Pasmanter et al. in this volume.

%%%%%%%%%%%%%%%%%%%%%%%%%%%%%%%%%%%%%%%%%%%%%%%%%%%%%%%%%%%%%%%%%%%%%%%%%
\section{Lagrangian Chaos}
\label{sec:3}

A problem of great interest concerns the
study of the spatial and temporal structure of the so-called passive fields,
indicating by this term passively quantities driven by the flow, such as 
 the temperature under certain conditions \cite{M83}.
 The equation for the evolution of a passive scalar field 
$\theta ({\bf x},t)$, advected by a velocity field ${\bf v}({\bf x},t)$, is 
\beq
\partial_t\theta +\nabla \cdot ({\bf v}\,\theta) =  \chi\,\nabla^2\theta
\label{tracer}
\eeq
where ${\bf v}({\bf x},t)$ is a given velocity field and 
$\chi$ is the molecular diffusion coefficient.

The problem (\ref{tracer}) can be studied through two
different approaches. Either one deals at any time with 
the field $\theta$ in the space domain covered by the fluid, or 
one deals with the trajectory of each fluid particle. The two approaches 
are usually designed as ``Eulerian''and ``Lagrangian'', although both of 
them are due to Euler \cite{L45}.
The two points of view are in principle equivalent.  

The motion of a fluid particle is determined by the differential
equation
\beq
{d{\bf x}\over d t}= {\bf v}({\bf x},t)
\label{lagr}
\eeq
which also describes the motion of test particles, for example a
powder embedded in the fluid, provided that the particles are neutral and
small enough not to perturb the velocity field, although large enough not 
to perform a Brownian motion. Particles of this type are commonly used 
for flow visualization in fluid mechanics experiments, see \cite{T88} and 
the contribution of  
Cenedese et al. in this volume.  
Let us note that the true equation for the motion of a material particle in
a fluid can be rather complicated \cite{MR83,CFPrV90}. 

It is now well established that even in regular velocity field
the motion of fluid particles can be very irregular \cite{H66,A84}.
In this case initially nearby trajectories diverge exponentially 
and one speaks of {\it Lagrangian chaos}.
In general, chaotic behaviors can arise in two-dimensional flow only 
for time dependent velocity fields in two dimensions, while it can
be present even for stationary velocity fields in three dimensions.

If $\chi=0$, it 
is easy to realize that (\ref{tracer}) is equivalent to (\ref{lagr}).
 In fact, we can write
\beq
\theta({\bf x},t)= \theta_o({T}^{-t} {\bf x})
\label{theta}
\eeq
where $\theta_o({\bf x})= \theta({\bf x}, t= 0)$ and ${T}$ is the formal
evolution operator of (\ref{lagr}) ,
\beq
{\bf x}(t)= {T}^t{\bf x}(0).
\label{eq:4}
\eeq

Taking into account the molecular diffusion $\chi$, 
(\ref{tracer}) is the Fokker-Planck equation of the Langevin 
 equation \cite{C43} 
\beq 
{d{\bf x} \over d t}= {\bf v}({\bf x}, t) + {\eta}(t)
\label{eq:5}
\eeq
where ${\eta}$ is a Gaussian process with zero mean and variance
\beq
\left\langle{\eta_i(t)\, \eta_j(t')}\right\rangle= 
 2\chi \delta_{ij}\,\delta(t-t').
\label{eq:6}
\eeq

In the following we will consider only incompressible flow
\beq
\nabla\cdot{\bf v}= 0
\label{eq:7}
\eeq
for which the dynamical system (\ref{lagr}) is conservative.   
In two dimensions, the constraint (\ref{eq:7}) is automatically 
satisfied assuming
\beq
v_1= {\partial \psi\over\partial x_2}, \qquad
v_2= -\,{\partial \psi\over\partial x_1}
\label{eq:8}
\eeq
where $\psi({\bf x},t)$ is the {\it stream function}.
Inserting (\ref{eq:8}) into (\ref{lagr}) the evolution equations 
become 
\beq
{ {d x_1} \over {d t} }= {\partial \psi\over\partial x_2}, \qquad
     { {d x_2} \over {d t} }= -\,{\partial \psi\over\partial x_1}.
\label{eq:9}
\eeq
Formally (\ref{eq:9}) is a Hamiltonian system with the Hamiltonian given by 
the stream function $\psi$.
 
%%%%%%%%%%%%%%%%%%%%%%%%%%%%%%%%%%%%%%%%%%%%%%%%%%%%%%%%%%%%
\subsection{Examples of Lagrangian chaos}
\label{sec:3.1}

As a first example we consider a $3d$ stationary 
velocity field, the so-called ABC flow
\beq 
{\bf v} = \left( A \sin z + C \cos y, \, B \sin x + 
    A \cos z, \, C \sin y + B \cos x \right)
\label{eq:11}
\eeq
where $A, \, B$ and $C$ are non zero real parameters. 
Because of the incompressibility condition, the evolution 
${\bf x} (0) \to {\bf x} (t)$
defines a volume preserving, dynamics. 

\begin{figure}[hbt]
\epsfxsize=12truecm
\centerline{\epsfbox{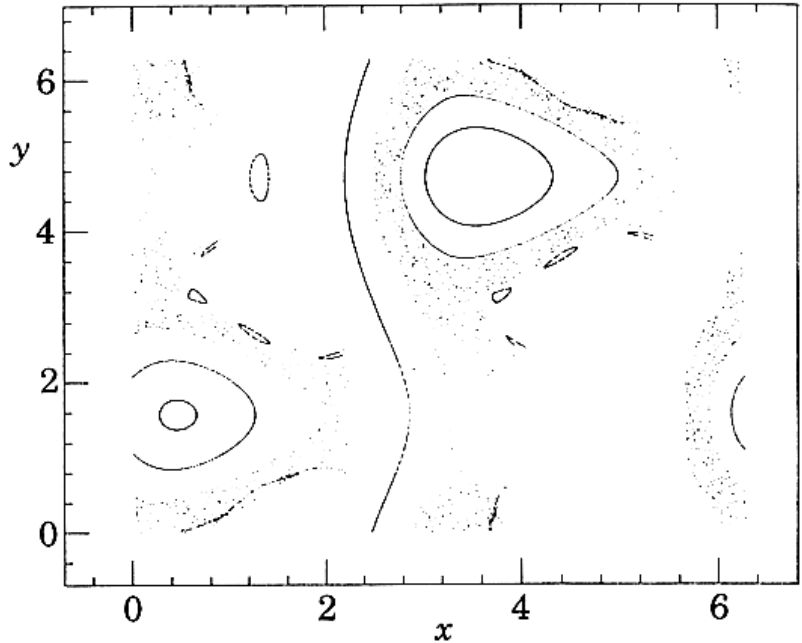}}
\caption{Intersections with the Poincar\'e section, plane $z=0$, 
of eight trajectories of the ABC flow with parameters $A=2.0$, $B=1.70$, 
$C=1.50$.
}
\label{fig:ABC}
\end{figure}

Arnold \cite{A65} 
argued that (\ref{eq:11}) is a good candidate for chaotic motion. 
Let us briefly repeat his elegant argument. For a steady state solution of the 
$3d$ Euler equation one has:
\begin{eqnarray}
%\begin{array}{lll}
\nabla \cdot {\bf v} &=& 0 \nonumber \\
{\bf v}\times(\nabla\times{\bf v}) &=& \nabla\alpha 
\label{eq:13} \\
\alpha &=& {P \over \rho} + {{\bf v}^2 \over 2} \nonumber 
%\end{array}
\end{eqnarray}
where $P$ is the pressure and $\rho$ the density. As a consequence of the 
Bernoulli theorem \cite{LL87},
 $\alpha( {\bf x} )$ is constant along a streamline -- 
that is a Lagrangian trajectory ${\bf x}(t)$.
One can easily verify that chaotic motion can appear only if 
$\alpha ({\bf x} )$ is constant (i.e. $\nabla \alpha ({\bf x}) = 0$) 
in a part of the space. Otherwise the trajectory would be confined on a 
$2d$ surface $\alpha ({\bf x})=$ constant, where the motion must be 
regular as a consequence of general arguments \cite{Ott93}.
In order to satisfy such a constraint, from (\ref{eq:13}) one has the 
Beltrami condition:
\beq
\nabla\times{\bf v}= \gamma({\bf x})\, {\bf v}.
\label{eq:14}
\eeq
The reader can easily verify that the field ${\bf v}$ given by (\ref{eq:11}) 
satisfy (\ref{eq:14}) (in this case $\gamma({\bf x}) =$ constant). 
Indeed, numerical experiments by H\'enon \cite{H66} provided evidence that 
Lagrangian motion under velocity (\ref{eq:11}) is chaotic
for typical values of the parameters $A$, $B$, and $C$
(see an example in Figure \ref{fig:ABC}).

In a two-dimensional incompressible stationary flows
the motion of fluid particles is given by a time independent 
Hamiltonian system with one 
degree of freedom and, since trajectories follow 
iso-$\psi$ lines, it is impossible to have chaos.
However, for explicit time dependent stream function $\psi$ the system 
(\ref{eq:11}) can exhibit chaotic motion \cite{Ott93}.  

In the particular case of time periodic velocity fields, 
${\bf v} ({\bf x}, t+T) = {\bf v} ({\bf x}, t)$, the trajectory
of (\ref{lagr}) can be studied in terms of 
discrete dynamical systems: the position ${\bf x}(t+T)$ is determined
in terms of ${\bf x}(t)$. The map  ${\bf x}(t) \to {\bf x}(t+T)$ 
will not depend on $t$ thus (\ref{lagr}) can be written in the form
\beq
{\bf x}(n+1) = {\bf F} [{\bf x}(n)] ,
\label{eq:18}
\eeq
where now the time is measured in units of the period $T$. Because of
incompressibility, the map (\ref{eq:18}) is conservative:
\beq
\bigl| \det {A} [{\bf x}] \bigr| = 1 ,  \qquad {\rm where}  
    \qquad   A_{i j} [{\bf x} ] =
 {\partial F_i [{\bf x}] \over \partial x_j }. 
\label{eq:19}
\eeq 
An  explicit deduction of the form of ${\bf F}$ for a general $2d$ or $3d$ flow 
is usually very difficult. However, in some simple model of
can be deduced on the basis of physical features \cite{AB86,CCTT87}.

%%%%%%%%%%%%%%%%%%%%%%%%%%%%%%%%%%%%%%%%%%%%%%%%%%%%%%%%%%%%%%%%%
\subsection{Eulerian properties and Lagrangian chaos}
\label{sec:3.2}

In principle, the evolution of the velocity field ${\bf v}$ is 
described by partial differential equations, e.g. Navier-Stokes or
Boussinesq equations.
However, often in weakly turbulent situations, a good approximation 
of the flow can be obtained by using a Galerkin approach, and reducing 
the Eulerian problem to a (small) system of $F$ ordinary
differential equations \cite{BF79,L87}. The motion of a fluid particle is
then described by the $(d+F)$-dimensional dynamical system
\beq
{d{\bf Q} \over d t}= {\bf f}({\bf Q},t)
\qquad
\hbox{\rm with}\ {\bf Q},\, {\bf f}\,\in {\rm I\!R}^F
\label{eq:31a}
\eeq
\beq
{d{\bf x}\over d t}= {\bf v}({\bf x},{\bf Q})  \qquad
\hbox{\rm with}\ {\bf x},\,{\bf v} \in {\rm I\!R}^d
\label{eq:31b}
\eeq
where $d$ is the space dimensionality and ${\bf Q}= (Q_1,...Q_F)$ are the 
$F$ variables, usually normal modes, which are a representation of
the velocity field ${\bf v}$. Note that the Eulerian equations (\ref{eq:31a})
do not depend on the Lagrangian part (\ref{eq:31b}) and can be solved 
independently.
 
In order to characterize the degree of chaos, three
different Lyapunov exponents can be defined \cite{FPV88}:
\begin{itemize}
\item{a)} $\lambda_{\rm E}$ for the Eulerian part (\ref{eq:31a}); 

\item{b)} $\lambda_{\rm L}$ for the Lagrangian part (\ref{eq:31b}), where
the evolution of the velocity field is assumed to be known;

\item{c)} $\lambda_{\rm T}$ per for the total system of the $d+F$ equations.
\end{itemize}
These Lyapunov exponents are defined as:
\beq
\lambda_{\rm E,L,T}= \lim_{t\to\infty}\,  {1 \over t}\,
           \ln{|{\bf z}(t)^{({\rm E,L,T})}| \over
               |{\bf z}(0)^{({\rm E,L,T})}|} 
\label{eq:32}
\eeq
where the evolution of the three tangent vectors ${\bf z}$ are given by the
linearized stability equations for the Eulerian part, for the Lagrangian
part and for the total system, respectively:
\beq
{d z_i^{({\rm E})}\over d t}= \sum_{j=1}^{F}
              \left.{\partial f_i\over \partial Q_j} \right|_{{\bf Q}(t)}\,
                          {z_j}^{({\rm E})}, \qquad 
                {\bf z}^{({\rm E})}\,\in {\rm I\!R}^F
\label{eq:33}
\eeq
\beq
{d z_i^{({\rm L})}\over d t}= \sum_{j=1}^{d}
              \left.{\partial v_i\over \partial x_j} \right|_{{\bf x}(t)}\,
                          {z_j}^{({\rm L})}, \qquad 
                {\bf z}^{({\rm L})}\,\in {\rm I\!R}^d
\label{eq:34}
\eeq
\beq
{d z_i^{({\rm T})}\over d t}= \sum_{j=1}^{d+F}
              \left.{\partial G_i\over \partial y_j} \right|_{{\bf y}(t)}\,
                          {z_j}^{({\rm T})}, \qquad 
                {\bf z}^{({\rm T})}\,\in {\rm I\!R}^{F+d}
\label{eq:35}
\eeq
and  ${\bf y}= (Q_1,\ldots,Q_F, x_1,\ldots,x_d)$ and 
${\bf G}=(f_1,\ldots,f_F,v_1,\ldots,v_d)$. 
The meaning of these Lyapunov exponents is evident:
\begin{itemize}
\item
{a)} $\lambda_{\rm E}$ is the mean exponential rate  of the increasing
 of the uncertainty in the knowledge of the velocity field (which is,
by definition, independent on the Lagrangian motion);

\item
{b)} $\lambda_{\rm L}$ estimates the rate at which the distance
$\delta x(t)$ between two fluid particles initially close increases with time,
when the velocity field is given, i.e. a particle pair in the same Eulerian
realization;

\item
{c)} $\lambda_{\rm T}$ is the rate of growth of the distance between
initially close particle pairs, when the velocity field is not known with
infinite precision.
\end{itemize}

There is no general relation between $\lambda_{\rm E}$ and $\lambda_{\rm L}$. 
One could expect that in presence of a chaotic velocity field the particle 
motion has to be chaotic. However, the inequality 
$\lambda_{\rm L} \ge \lambda_{\rm E}$ -- even if generic -- 
sometimes does not hold, e.g. 
in some systems like the Lorenz model \cite{FPV88} 
and in generic $2d$ flows when the Lagrangian motion happens around 
well defined vortex structures \cite{BBPrV94} as discussed in 
the following.
On the contrary, one has \cite{CFPV91} 
\beq
\lambda_{\rm T} = {\rm max}\, (\lambda_{\rm E},\lambda_{\rm L}).
\label{eq:36}
\eeq

%%%%%%%%%%%%%%%%%%%%%%%%%%%%%%%%%%%%%%%%%%%%%%%%%%%%%%%%%%%%%%%%
\subsection{Lagrangian chaos in two dimensional flows}
\label{sec:3.3}

Let us now consider the two--dimensional Navier-Stokes equations 
with periodic boundary conditions at low Reynolds numbers, for
which we can expand the stream function $\psi$ in Fourier series 
and takes into account only the first $F$ terms
\cite{BF79,L87},
\beq 
\psi=-i \sum_{j=1}^{F}\  k_j^{-1} 
        Q_{_j} e^{i {\bf k}_j {\bf x} } + {\rm c.  c.} \, ,
\label{eq:37}
\eeq
where c.c. indicates the complex conjugate term
and ${\bf Q}= (Q_1,\ldots,Q_F)$ 
are the Fourier coefficients.
Inserting (\ref{eq:37}) into the Navier-Stokes equations and by an 
appropriate time rescaling, we obtain the system of $F$ ordinary 
differential equations
\beq
{d Q_j \over d t} = - k_j^2\, Q_j +
         \sum_{l,m} A_{j l m} Q_l Q_m + f_j,
\label{eq:38}
\eeq
in which $f_j$ represents an external forcing.

Franceschini and coworkers have studied this truncated model with $F=5$ 
and $F=7$ \cite{BF79,L87}.
The forcing were restricted to the $3^{th}$ mode
$f_j={\rm Re}\,\delta_{j,3}$ \cite{L87}.
For $F=5$ and ${\rm Re} < {\rm Re}_1 = 22.85\ldots$, there are four stable stationary 
solutions, say $\widehat{\bf Q}$, and $\lambda_{\rm E} < 0$. At 
${\rm Re} = {\rm Re}_1$, these solutions become unstable, via a Hopf 
bifurcation \cite{MM75}, and four stable periodic 
orbits appear, still implying $\lambda_{\rm E}=0$. 
For ${\rm Re}_1<{\rm Re}< {\rm Re}_2= 28.41\ldots$,
one thus finds the stable limit cycles:
\beq
{\bf Q} (t)= \widehat{\bf Q} + ({\rm Re}-{\rm Re}_1)^{1/2} \delta{\bf Q}(t)
              + O({\rm Re}-{\rm Re}_1)  
\label{eq:39}
\eeq
where $\delta{\bf Q}(t)$ is periodic with period
\beq
T({\rm Re})= T_0 + O({\rm Re}-{\rm Re}_1) \qquad  T_0 = 0.7328\ldots
\label{eq:40}
\eeq 
At ${\rm Re}={\rm Re}_2$, these limit cycles lose stability and there is a 
period doubling cascade toward Eulerian chaos.

\begin{figure}[hbt]
\epsfxsize=12truecm
\centerline{\epsfbox{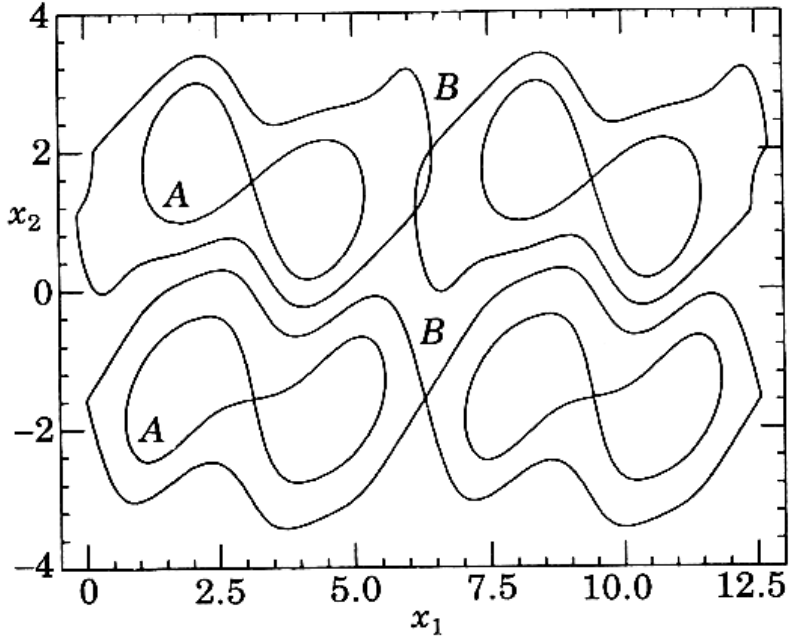}}
\caption{Structure of the separatrices in the 5-mode model 
(\ref{eq:37}) with ${\rm Re} = {\rm Re}_1 - 0.05$. 
}
\label{fig:separatrices}
\end{figure}

Let us now  discuss the Lagrangian behavior of a fluid particle.
For ${\rm Re} < {\rm Re}_1$, the stream function is asymptotically stationary, 
$\psi({\bf x},t) \to \widehat{\psi}({\bf x})$, and the corresponding 
one-dimensional Hamiltonian is time-independent,  
therefore Lagrangian trajectories are regular.
For ${\rm Re}={\rm Re}_1+\epsilon$ the stream function becomes time 
dependent 
\beq
\psi({\bf x},t)= \widehat{\psi}({\bf x}) +
                  \sqrt{\epsilon}\,\delta\psi({\bf x},t) +
                  O(\epsilon),
\label{eq:41}
\eeq
where $\widehat{\psi}({\bf x})$ is given by $\widehat{{\bf Q}}$
and $\delta \psi$ is periodic in ${\bf x}$ and in $t$ with period $T$.
The region of phase space, 
here the real two-dimensional space, adjacent to a separatrix is very sensitive 
to perturbations, even of very weak intensity. 
Figure \ref{fig:separatrices} shows the structure of the separatrices, 
i.e. the orbits of infinite periods at ${\rm Re}={\rm Re}_1 - 0.05$. 

\begin{figure}[hbt]
\epsfxsize=12truecm
\centerline{\epsfbox{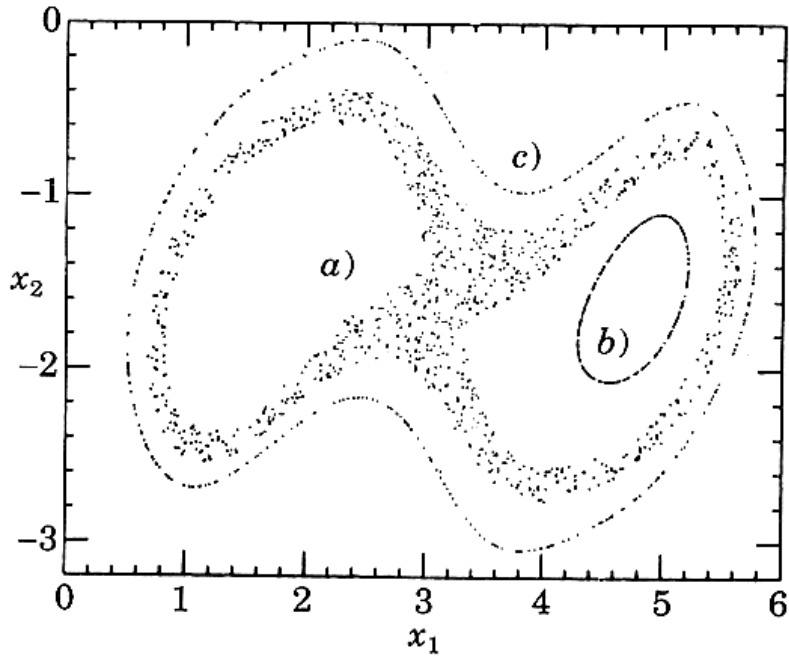}}
\caption{Poincar\'e map for three trajectories of the 5-mode model 
with ${\rm Re} = {\rm Re}_1 + 0.05$. The initial conditions are selected 
close to a separatrix, case a) ($x_1(0) = 3.2$, $x_2(0) = -1.6$), 
or far from the separatrices, cases b) ($x_1(0)=4.3$, $x_2(0)=-2.0$) 
and c) ($x_1(0)=4.267$, $x_2(0)=-3.009$). 
}
\label{fig:map}
\end{figure}

Indeed, generically in 
one-dimensional Hamiltonian systems, a periodic perturbation gives origin to
stochastic layers around the separatrices where the motion is chaotic, as
consequence of unfolding and crossing of the stable and unstable manifolds 
in domains centered at the hyperbolic fixed points \cite{chirikov79,Ott93}.
One has strong  
numerical evidence for the existence of the chaotic regions, see Figure 
\ref{fig:map}.  

Chaotic and regular motion for small $\epsilon={\rm Re}_1-{\rm Re}$ 
can be studied by the Poincar\'e map 
\beq
{\bf x}(nT)\to{\bf x}(nT + T). 
\label{eq:42}
\eeq
The period $T(\epsilon)$ is computed numerically.
 The size of the stochastic 
layers rapidly increase with $\epsilon$. At 
$\epsilon=\epsilon_{\rm c} \approx 0.7$ they 
overlap and it is practically impossible to distinguish between regular and 
chaotic zones. 
At $\epsilon > \epsilon_{\rm c}$ 
there is always diffusive motion.

We stress that this scenario for the onset of Lagrangian chaos in 
two-dimensional fluids
is generic and does not depend on the particular truncated
model. In fact, it is only related to the appearance of stochastic layers 
under the effects of small time-dependent perturbations in one-dimensional
integrable Hamiltonian systems. As consequence of a general features of
one-dimensional Hamiltonian systems we expect that a stationary stream
function becomes time periodic through a Hopf bifurcation as occurs 
for all known truncated models of Navier-Stokes equations. 

We have seen that there is no simple relation between Eulerian 
and Lagrangian behaviors. In the following, we shall discuss two 
important points:
\begin{itemize}
\item
{(i)} what are the effects on the Lagrangian chaos of  the transition
to Eulerian chaos, i.e. from $\lambda_{\rm E}=0$ to $\lambda_{\rm E}>0$.

\item
{(ii)} whether a chaotic velocity field ($\lambda_{\rm E}>0$)
always implies an erratic motion of fluid particles. 
\end{itemize}

\begin{figure}[hbt]
\epsfxsize=12truecm
\centerline{\epsfbox{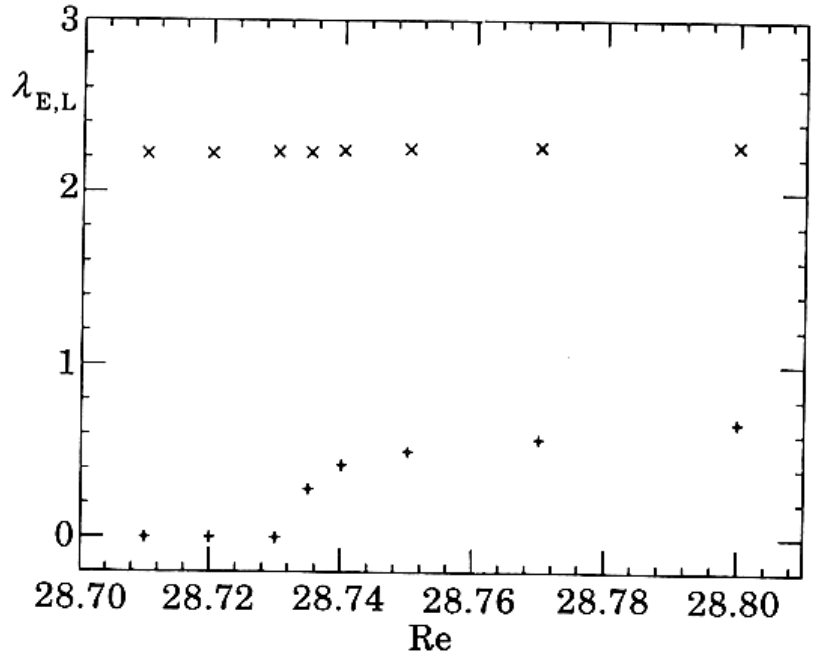}}
\caption{Lyapunov exponents $\lambda_E$ ($+$) and $\lambda_L$ 
($\times$) as  
function of ${\rm Re}$ around ${\rm Re}_c$, for the 5-mode model.
}
\label{fig:lyapexp}
\end{figure}

The first point can be studied again within the $F=5$ modes
model (\ref{eq:38}).
Increasing ${\rm Re}$, the limit cycles bifurcate to new double 
period orbits followed by a period doubling transition to chaos 
and a strange attractor appears at ${\rm Re}_c \approx 28.73$,
where $\lambda_{\rm E}$ becomes positive. 
These transitions have no signature on Lagrangian behavior,
as it is shown in Figure \ref{fig:lyapexp}, 
i.e. the onset of Eulerian chaos has
no influence on Lagrangian properties.

This feature should be valid in most situations, since  it is natural 
to expect that in generic cases there is a strong separation of the
characteristic times for Eulerian and Lagrangian behaviors.

The second point -- the conjecture that a chaotic velocity field 
always implies chaotic motion of particles -- looks very reasonable.
Indeed, it appears to hold in many systems \cite{CFPV91}.
Nevertheless, one can find a class of systems where it is false, 
e.g. the equations (\ref{eq:31a}), (\ref{eq:31b}) may exhibit 
Eulerian chaoticity $\lambda_{\rm E} > 0$, even if $\lambda_{\rm L}=0$ 
\cite{BBPrV94}.

Consider for example the motion of $N$ point vortices in the plane 
with circulations $ \Gamma_i$ and positions $(x_i(t),y_i(t))$
($i=1,..N$) \cite{A83}:
\beq
\Gamma_i {{d x_i } \over {d t}} =
{{ \partial H }\over {\partial y_i}}
\label{eq:49}
\eeq
\beq
\Gamma_i {{d y_i } \over {d t}} =
- {{ \partial H }\over {\partial x_i}}
\label{eq:50}
\eeq
where
\beq
H = -{1 \over {4 \pi} } \sum_{i \ne j} \Gamma_i \Gamma_j 
\ln r_{ij}
\label{eq:51}
\eeq
and $r_{ij}^2 = (x_i-x_j)^2+(y_i-y_j)^2$. 

The motion of $N$ point vortices is described in an Eulerian 
phase space with $2N$ dimensions. Because of the presence of 
global conserved quantities, a system of three vortices is 
integrable and there is no exponential divergence of nearby 
trajectories in phase space.
For $N \ge 4$, apart from non generic initial conditions and/or values of the 
parameters $\Gamma_i$, the system is chaotic \cite{A83}.

The motion of a passively advected particle 
located in $(x(t),y(t))$ in the velocity field defined by 
(\ref{eq:49}-\ref{eq:50}) is given
\beq
         {{d x } \over {d t}} =
   - \sum_i {{\Gamma_i} \over {2 \pi}} {{y-y_i} \over {R_i^2}} 
\label{eq:52}
\eeq
\beq
         {{d y } \over {d t}} =
    \sum_i {{\Gamma_i} \over {2 \pi}} {{x-x_i} \over {R_i^2}} 
\label{eq:53}
\eeq
where $R_i^2 = (x-x_i)^2+(y-y_i)^2$.

Let us first consider the motion of advected particles in a three-vortices
(integrable) system in which $\lambda_E=0$.
In this case, the stream function for the advected 
particle is periodic in time and the expectation is that the advected
particles may display chaotic behavior. 
The typical 
trajectories of passive particles which have initially been placed 
respectively in close proximity of a 
vortex center or in the background field between the vortices
display a very different behavior. The particle
seeded close to the vortex center displays a regular motion 
around the vortex and thus $\lambda_L=0$;
by contrast, the particle in the background field 
undergoes an irregular and aperiodic trajectory, and $\lambda_L$ is
positive.

\begin{figure}[hbt]
\epsfxsize=8truecm
\centerline{\epsfxsize=8truecm,\epsfbox{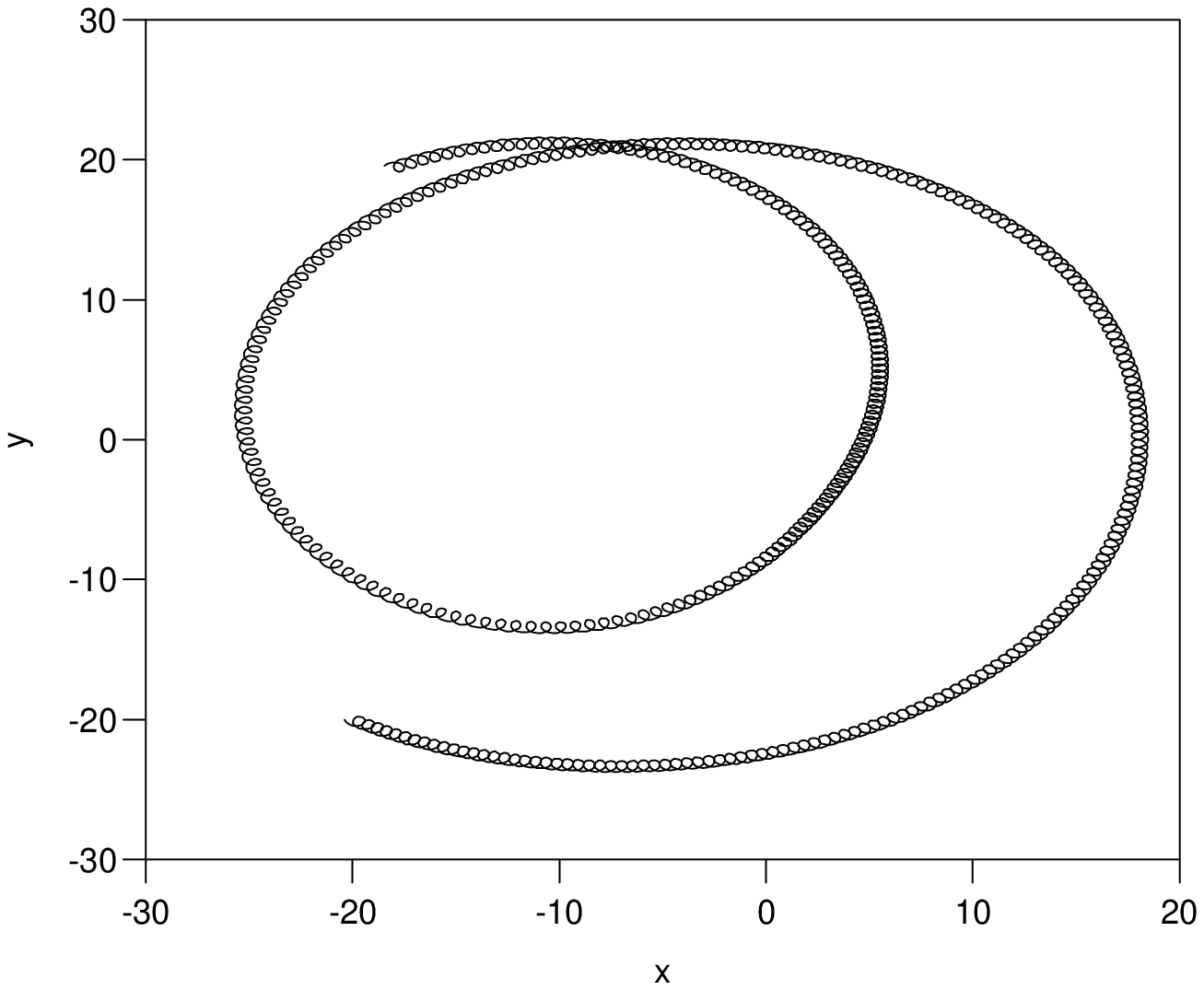},\epsfxsize=8truecm,\epsfbox{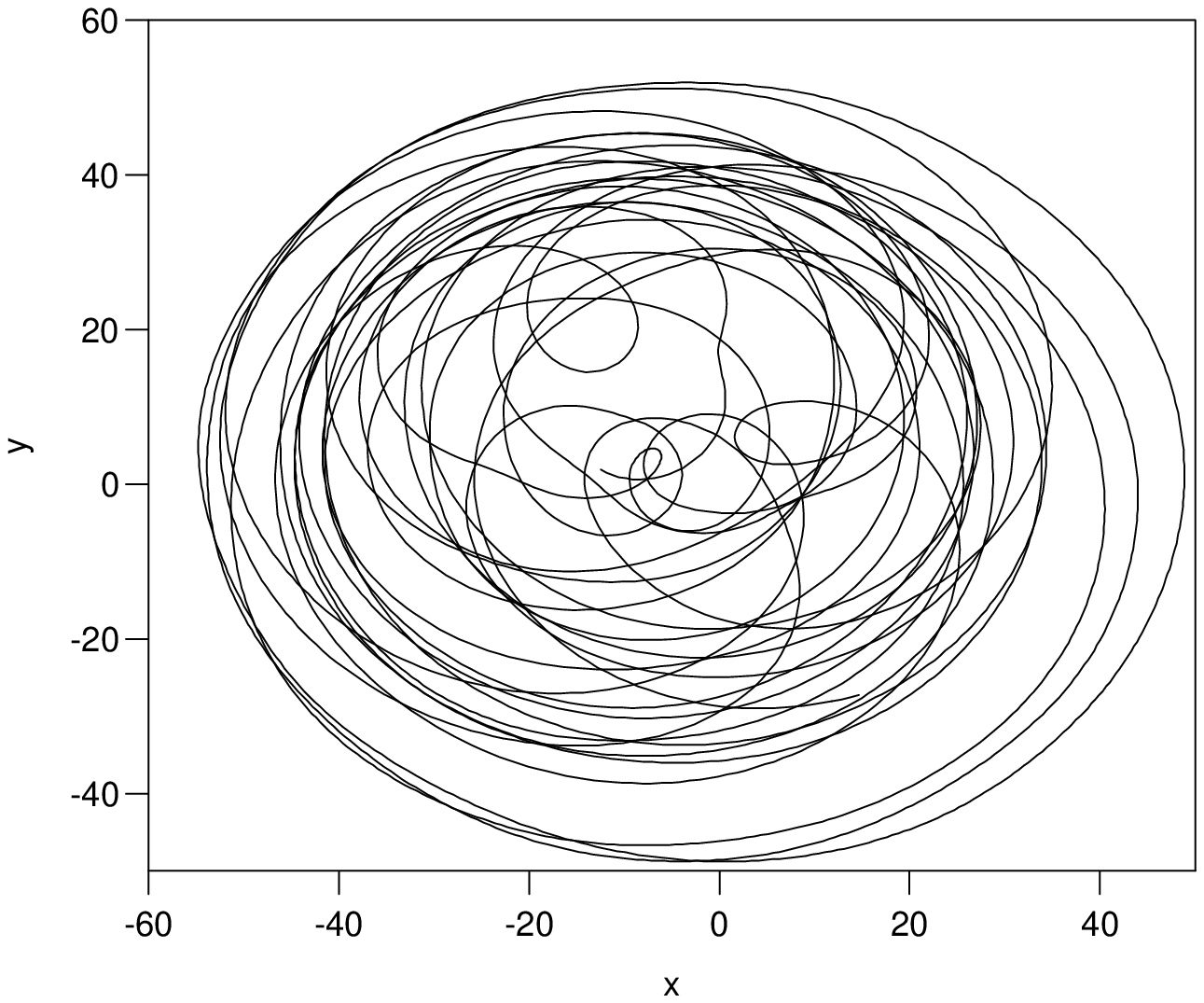}}
\caption{Particle trajectories in the four-vortex system.
Eulerian dynamics in this case is chaotic. The left panel
shows a regular Lagrangian trajectory while the right panel
shows a chaotic Lagrangian trajectory. The different behavior
of the two particles is due to different initial conditions.
}
\label{fig:bohr}
\end{figure}

We now discuss a case where the Eulerian flow is chaotic i.e. $N=4$
point vortices. Let us consider again the trajectory of a passive 
particle deployed in proximity of a vortex center.
As before, the particle rotates around the moving 
vortex. The vortex motion is chaotic; consequently, the particle position 
is unpredictable on large times as is the vortex position. 
Nevertheless, the Lagrangian Lyapunov exponent for this trajectory is zero
(i.e. two initially close particles around the vortex remain close),
even if the Eulerian Lyapunov exponent is positive, see 
Figure \ref{fig:bohr}.

This result indicates once more that there is no strict link between 
Eulerian and Lagrangian chaoticity.  

One may wonder whether a much more complex Eulerian flow, such as 
$2d$ turbulence, may give the same scenario for particle advection: i.e.
regular trajectories close to the vortices and chaotic behavior 
between the vortices.
It has been shown that this is indeed the case \cite{BBPrV94} and that the
chaotic nature of the trajectories of advected particles is not strictly 
determined by the complex time evolution of the turbulent flow.

We have seen that there is no general relation between 
Lagrangian and Eulerian chaos. In the typical situation Lagrangian chaos 
may appear also for regular velocity fields. However, 
it is also possible to have the opposite situation, with 
$\lambda_{\rm L}=0$ in presence of Eulerian chaos, as in the example
of Lagrangian motion inside vortex structures.
As an important consequence of this discussion we remark that 
it is not possible to separate Lagrangian and Eulerian properties 
in a measured trajectory, e.g. a buoy in the oceanic currents 
 \cite{OKPB86}. Indeed, using the standard methods for
data analysis \cite{GP83}, from Lagrangian
trajectories one extracts the total Lyapunov exponent 
$\lambda_{\rm T}$ and not $\lambda_{\rm L}$ or $\lambda_{\rm E}$.

%%%%%%%%%%%%%%%%%%%%%%%%%%%%%%%%%%%%%%%%%%%%%%%%%%%%%%%%%%%%%%%%%%%%
%%%%%%%%%%%%%%%%%%%%%%%%%%%%%%%%%%%%%%%%%%%%%%%%%%%%%%%%%%%%%%%%%%%%
\section{Transport and diffusion}
\label{sec:4}

The simplest model of diffusion is the Brownian motion, the
erratic movement of a grains suspended in liquid observed by the botanist
Robert Brown as early as in 1827.
After the fundamental work of Einstein \cite{einstein05} and 
Langevin \cite{langevin08}, Brownian motion become the prototypical
example of stochastic process.

%%%%%%%%%%%%%%%%%%%%%%%%%%%%%%%%%%%%%%%%%%%%%%%%%%%%%%%%%%%%%%%%%%%%
\subsection{The random walk model for Brownian motion}
\label{sec:4.1}

In order to study more in detail the properties of diffusion,
let us introduce the simplest model of Brownian motion, i.e.
the one-dimensional random walk. The walker moves on a line
making discrete jumps $v_i=\pm 1$ at discrete times. The position
of the walker, started at the origin at $t=0$, will be
\begin{equation}
R(t) = \sum_{i=1}^{t} v_i
\label{eq:4.8}
\end{equation}
Assuming equiprobability for left and right jumps (no mean motion),
the probability that at time $t$ the walker is in position $x$ will
be
\begin{equation}
p_{t}(R=x) = prob\left[
\begin{array}{c}
{t-x \over 2} steps -1 \\
{t+x \over 2} steps +1
\end{array}
\right] = {1 \over 2^{t}} 
\left(
\begin{array}{c}
t \\
{t+x \over 2}
\end{array}
\right)
\label{eq:4.9}
\end{equation}
For large $t$ and $x$ (i.e. after many microscopic steps) we can use
Stirling approximation and get
\begin{equation}
p_{t}(x) = \sqrt{{t \over 2 \pi(t^2-x^2)}} \exp{\left[
- {t+x \over 2} \ln {t+x \over 2} - {t-x \over 2} \ln {t-x \over 2}\right]}
\label{eq:4.10}
\end{equation}
The core of the distribution recovers the well known Gaussian 
form, i.e. for $x \ll t$ from (\ref{eq:4.10}) we get
\begin{equation}
p_{t}(x) = \sqrt{{1 \over 2 \pi t}} \exp \left(-{x^2 \over 2 t}\right)
\label{eq:4.11}
\end{equation}

From (\ref{eq:4.11}) one obtains that the variance of the
displacement follows diffusive behavior, i.e.
\begin{equation}
\langle x^2(t) \rangle = 2 t
\label{eq:11bis}
\end{equation}
We stress that diffusion is obtained only asymptotically 
(i.e. for $t \to \infty$).
This is a consequence of central limit theorem which assures 
Gaussian distributions and diffusive behavior in the limit of many
independent jumps.
The necessary, and sufficient, condition for observing diffusive regime
is the existence of a finite correlation time (here represented by
discrete time between jumps) for the microscopic dynamics. 
Let us stress that this is the important ingredient for diffusion, and
{\it not} a stochastic microscopic dynamics. We will see below that
diffusion can arise even in completely deterministic systems.

\begin{figure}[hbt]
\epsfxsize=12truecm
\centerline{\epsfbox{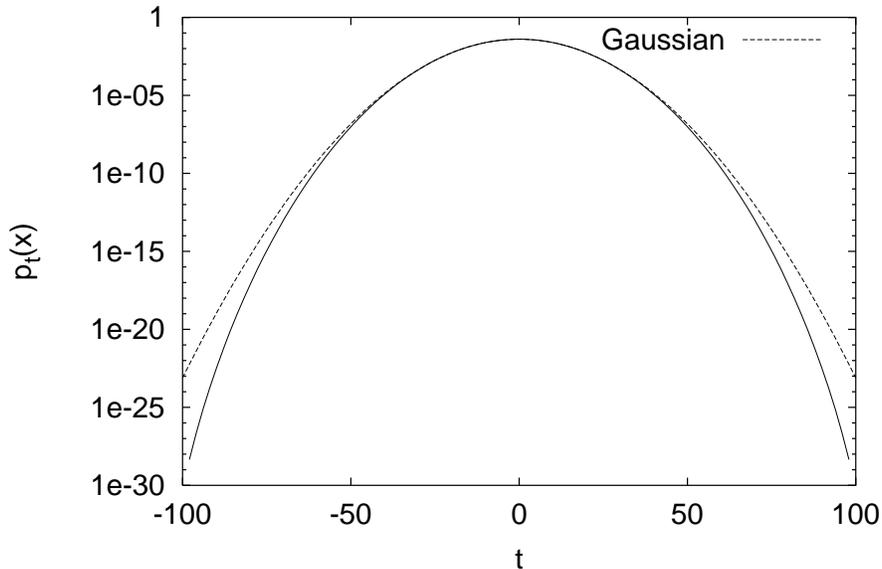}}
\caption{Probability distribution function of the one-dimensional
random walk after $t=100$ steps. The dashed line is the 
Gaussian distribution}
\label{fig1}
\end{figure}

Another important remark is that Gaussian distribution (\ref{eq:4.11}) is
intrinsic of diffusion process, independent on the distribution of 
microscopic jumps: indeed only the first two moments of $v_i$ enter
into expression (\ref{eq:4.11}).
This is, of course, the essence of the
central limit theorem. Following the above derivation, it is clear that
this is true only in the core of the distribution. The far tails 
keep memory of the microscopic process and are, in general, 
not Gaussian.
As an example, in Figure~\ref{fig1} we plot the pdf $p_{t}(x)$ at 
step $t=100$ compared with the Gaussian approximation. Deviations
are evident in the tails.

The Gaussian distribution (\ref{eq:4.11}) can be obtained as the solution
of the diffusion equation which governs the evolution of the
probability in time. This is the Fokker-Planck equation for the
particular stochastic process. A direct way to relate the 
one-dimensional random walk to the diffusion equation is obtained
by introducing the master equation, i.e. the time evolution of the
probability \cite{G85}:
\begin{equation}
p_{t+1}(x) = {1 \over 2} p_{t}(x-1) + {1 \over 2} p_{t}(x+1) \, .
\label{eq:4.12}
\end{equation}
In order to get a continuous limit, we introduce explicitly 
the steps $\Delta x$ and $\Delta t$ and write
\begin{equation}
{p_{t+\Delta t}(x)-p_{t}(x) \over \Delta t} = 
{(\Delta x)^2 \over 2 \Delta t}
{p_{t}(x+\Delta x)+p_{t}(x-\Delta x)-2 p_{t}(x) \over 
(\Delta x)^2} \, .
\label{eq:4.13}
\end{equation}
Now, taking the limit $\Delta x, \Delta t \to 0$ in such a way
that $(\Delta x)^2/\Delta t \to 2 D$ (the factor $2$ is purely conventional)
we obtain the diffusion equation
\begin{equation}
{\partial p(x,t) \over \partial t} = D {\partial^2 p(x,t) \over 
\partial x^2} \, .
\label{eq:4.14}
\end{equation}
The way the limit $\Delta x, \Delta t \to 0$ is taken reflects the 
scaling invariance property of diffusion equation.
The solution to (\ref{eq:4.14}) is readily obtained as
\begin{equation}
p(x,t) = {1 \over \sqrt{4 \pi D t}} \exp \left(-{x^2 \over 4 D t}\right) \, .
\label{eq:4.15}
\end{equation}

Diffusion equation (\ref{eq:4.14}) is here written for the 
probability $p(x,t)$ of observing a marked particle (a tracer)
in position $x$ at time $t$. 
The same equation can have another interpretation, in which 
$p(x,t)=\theta(x,t)$ represents the concentration of a scalar quantity 
(marked fluid, temperature, pollutant) as
function of time. The only difference is, of course, in the
normalization. 

%%%%%%%%%%%%%%%%%%%%%%%%%%%%%%%%%%%%%%%%%%%%%%%%%%%%%%%%%%%%%%%%%%%%
\subsection{Less simple transport processes}
\label{sec:4.2}

As already stated time decorrelation is the key ingredient for diffusion.
In the random walker model it is a consequence of randomness:
the steps $v_i$ are random uncorrelated variables and this assures 
the applicability of central limit theorem. 
But we can have a finite time correlation and thus diffusion also without 
randomness. To be more specific,
let us consider the following deterministic model 
(standard map \cite{chirikov79}):
\begin{equation}
\left\{
\begin{array}{ccc}
J(t+1) &=& J(t) + K \sin \theta(t) \\
\theta(t+1) &=& \theta(t)+J(t+1) 
\end{array}
\right.
\label{eq:4.16}
\end{equation}

\begin{figure}[htb]
\epsfxsize=12truecm
\centerline{\epsfbox{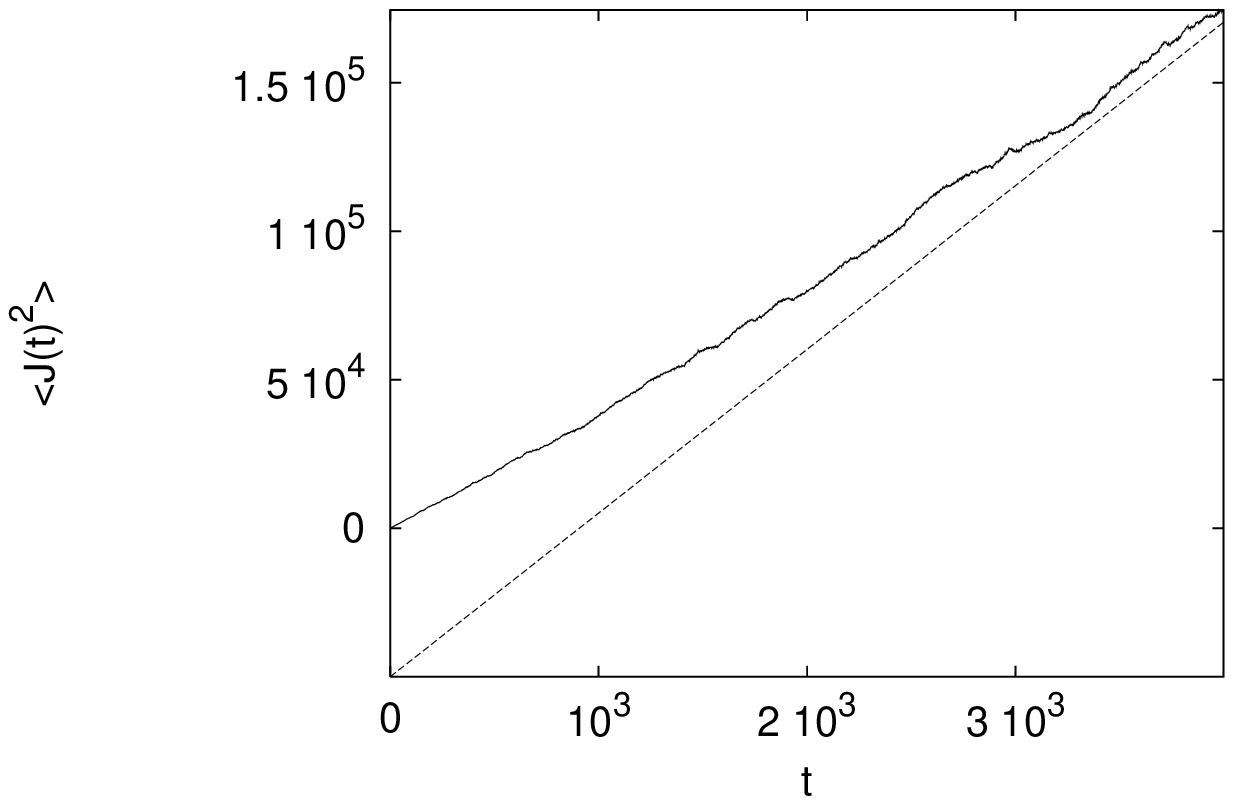}}
\caption{Square dispersion $\langle J(t)^2 \rangle$ for 
the standard map at $K=10.5$. The dashed line is the 
RPA prediction.}
\label{fig2}
\end{figure}

The map is known to display large-scale 
chaotic behavior for $K>K_c \simeq 0.9716$ and, as a consequence of 
deterministic chaos, $J(t)$ has diffusive behavior.  
For large times, $J(t)$ is large and thus the angle $\theta(t)$
rotates rapidly. In this limit, we can assume that at each step
$\theta(t)$ decorrelates and thus write
\begin{equation}
J(t)^2 = K^2 \left(\sum_{t'=1}^{t} \sin \theta(t') \right)^2
\simeq K^2 \langle \sin^2 \theta \rangle t = 2 D t
\label{eq:4.17}
\end{equation}
The diffusion coefficient $D$, in the random phase
approximation, i.e. assuming that $sin \theta(t)$ is not correlated 
with $sin \theta(t')$ for $t \neq t'$,
is obtained by the above expression as $D_{RPA}=K^2/4$.
In Figure~\ref{fig2} we plot a numerical simulation obtained 
with the standard map. Diffusive behavior is clearly visible 
at long time.

The two examples discussed above are in completely different classes: 
stochastic for the random walk (\ref{eq:4.8}) and deterministic for the 
standard map (\ref{eq:4.16}).
Despite this difference in the microscopic dynamics, both lead 
to a macroscopic diffusion equation and Gaussian distribution. 
This demonstrates how diffusion equation is of general applicability.

%%%%%%%%%%%%%%%%%%%%%%%%%%%%%%%%%%%%%%%%%%%%%%%%%%%%%%%%%%%%%%%%%%%%
\subsection{Advection--diffusion}
\label{sec:4.3}

Let us now consider the more complex situation of 
dispersion in a non-steady fluid with velocity field 
${\bf v}({\bf x},t)$.
For simplicity will we consider incompressible flow 
(i.e. for which ${\bf \nabla } \cdot {\bf v}=0$)
which can be laminar or turbulent, solution of Navier-Stokes
equation or synthetically generated according to a given algorithm.
In presence of ${\bf v}({\bf x},t)$, the diffusion equation (\ref{eq:4.14})
becomes the {\it advection-diffusion} equation for the concentration  
$\theta({\bf x},t)$ (\ref{tracer}).
This equation is linear in $\theta$ but nevertheless
it can display very interesting and non trivial properties
even in presence of simple velocity fields, as a consequence of Lagrangian 
chaos.
In the following we will consider a very simple example of 
diffusion in presence of an array of vortices. The example 
will illustrate in a nice way the basic mechanisms
and effects of interaction between deterministic 
(${\bf v}$) and stochastic ($D$) components.

Let us remark that we will not consider here the problem of transport
in turbulent velocity field. This is a very classical problem,
with obvious and important applications, which has recently attracted
a renewal theoretical interest as a model for understanding the
basic properties of turbulence \cite{SS00}.

Before going into the example, let us make some general consideration.
We have seen that in physical systems the molecular diffusivity
is typically very small. Thus in (\ref{tracer}) the advection term
dominates over diffusion. This is quantified by the Peclet number, which
is the ratio of the typical value of the advection term to the 
diffusive term
\begin{equation}
Pe = {v_0 l_0 \over D}
\label{eq:2.2}
\end{equation}
where $v_{0}$ is the typical velocity at the typical scale of the flow
$l_{0}$. With $\tau_{0} \simeq l_0/v_0$ we will denote the typical
correlation time of the velocity. 

The central point in the following discussion is the concept of
{\it eddy diffusivity}. The idea is rather simple and dates back to 
the classical work of Taylor \cite{T21}.
To illustrate this concept, let us consider a Lagrangian description 
of dispersion in which the trajectory of a tracer ${\bf x}(t)$ is given by 
(\ref{lagr}).
Being interested in the limit $Pe \to \infty$, in the following we 
will neglect, just for simplicity, 
the molecular diffusivity $D$, which is generally much 
lesser that the effective dynamical diffusion coefficient.

Starting from the origin, ${\bf x}(0)=0$, and assuming
$\langle {\bf v} \rangle=0$ we have 
$\langle {\bf x}(t)\rangle=0$ for ever. The square displacement,
on the other hand, grows according to
\begin{equation}
{d \over d t} \langle {1 \over 2} {\bf x}(t)^2 \rangle =
\langle {\bf x}(t) \cdot {\bf v}_{L}(t) \rangle =
\int_{0}^{t} \langle {\bf v}_{L}(s) \cdot {\bf v}_{L}(t) \rangle ds
\label{eq:2.4}
\end{equation}
where we have introduced, for simplicity of notation,
the Lagrangian velocity ${\bf v}_{L}(t)={\bf v}({\bf x}(t),t)$.
Define the Lagrangian correlation time $\tau_{L}$ from 
\begin{equation}
\int_{0}^{\infty} \langle {\bf v}_{L}(s) \cdot {\bf v}_{L}(0) \rangle ds
= \langle {\bf v}_{L}(0)^2 \rangle \tau_{L}
\label{eq:2.5}
\end{equation}
and assume that the integral converge so that $\tau_{L}$ is finite. 
From (\ref{eq:2.4}), for $t \gg \tau_{L}$ we get
\begin{equation}
\langle {\bf x}(t)^2 \rangle  = 2 \tau_{L} \langle {\bf v}_{L}^2 \rangle t
\label{eq:2.6}
\end{equation}
i.e. diffusive behavior with diffusion coefficient (eddy diffusivity)
$D^{E}=\tau_{L} \langle {\bf v}_{L}^2 \rangle$.

\begin{figure}[hbt]
\centerline{
\setlength{\unitlength}{2mm}
\begin{picture}(40,10)(40,10)
\put(30,0){\framebox(50,15)}
\put(40,0){\line(0,1){15}}
\put(50,0){\line(0,1){15}}
\put(60,0){\line(0,1){15}}
\put(70,0){\line(0,1){15}}
\put(35,7.5){\oval(6,9)}
\put(45,7.5){\oval(6,9)}
\put(55,7.5){\oval(6,9)}
\put(65,7.5){\oval(6,9)}
\put(75,7.5){\oval(6,9)}
\put(32,8){\vector(0,1){0}}
\put(38,7){\vector(0,-1){0}}
\put(42,7){\vector(0,-1){0}}
\put(48,8){\vector(0,1){0}}
\put(52,8){\vector(0,1){0}}
\put(58,7){\vector(0,-1){0}}
\put(62,7){\vector(0,-1){0}}
\put(68,8){\vector(0,1){0}}
\put(72,8){\vector(0,1){0}}
\put(78,7){\vector(0,-1){0}}
\put(45,-2){\vector(1,0){5}}
\put(45,-2){\vector(-1,0){5}}
\put(45,-4){$l_0$}
\put(50,16){\vector(1,0){1}}
\put(50,16){\vector(-1,0){1}}
\put(50,17){$\delta$}
\end{picture}
}
%\centerline{\epsfbox{vort1.ps}}
\vspace{2.5cm}
\caption{Cellular flow model. $l_{0}$ is the size of vortices,
$\delta$ is the thickness of the boundary layer.}
\label{fig3}
\end{figure}
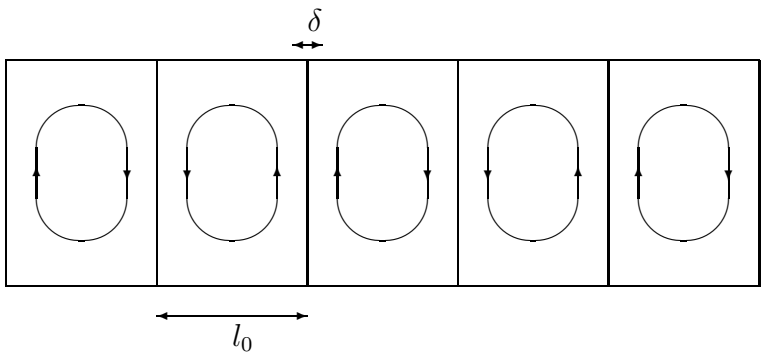

This simple derivation shows, once more, that diffusion has to be 
expected in general in presence of a finite correlation time $\tau_{L}$.
Coming back to the advection-diffusion equation (\ref{tracer}), 
the above argument means that for $t \gg \tau_{L}$ we expect
that the evolution of the concentration, for scales larger than $l_0$,
can be described by an effective diffusion equation, i.e.
\begin{equation}
{\partial \langle \theta \rangle \over \partial t} 
= D^{E}_{ij} {\partial^2 \langle \theta \rangle \over 
\partial x_i \partial x_j}
\label{eq:2.7}
\end{equation}
The computation of the eddy diffusivity for a given Eulerian 
flow is not an easy task. It can be done explicitly only in the 
case of simple flows, for example by means of homogenization theory
\cite{MK99,BCVV95}. In the general case it is relatively simple 
\cite{BCVV95} to
give some bounds, the simplest one being $D^{E} \ge D$,
i.e. the presence of a (incompressible)
velocity field enhances large-scale transport.
To be more specific, let us now consider the example of transport 
in a one-dimensional
array of vortices (cellular flow) sketched in Figure~\ref{fig3}.
This simple two-dimensional flow is useful for illustrating the
transport across barrier. Moreover, it naturally arises in several
fluid dynamics contexts, such as, for example, convective patterns 
\cite{SG88}.

%%%%%%%%%%%%%%%%%%%%%%%%%%%%%%%%%%%%%%

Let us denote by $v_0$ the typical velocity inside the cell of
size $l_0$ and let $D$ the molecular diffusivity. Because of the 
cellular structure, particles inside a vortex can exit only as a 
consequence of molecular diffusion. In a characteristic vortex time
$\tau_0 \sim  l_0 /v_0$, only the particles in the boundary layer
of thickness $\delta$ can cross the separatrix where
\begin{equation}
\delta^2 \sim D \tau_0 \sim D {l_0 \over v_0} \, .
\label{eq:2.8}
\end{equation}
These particles are ballistically advected by the velocity field 
across the vortex so they see a ``diffusion coefficient'' $l_0^2/\tau_0$.
Taking into account that this fraction of particles is $\delta/l_0$
we obtain an estimation for the effective diffusion coefficient as
\begin{equation}
D^{E} \sim {\delta \over l_0} {l_0^2 \over \tau_0} \sim 
\sqrt{D l_0 v_0} \sim D Pe^{1/2}
\label{eq:2.9}
\end{equation}

The above result, which can be made more rigorous, was confirmed by
nice experiments made by Solomon and Gollub \cite{SG88}. 
Because, as already stressed above, typically $Pe \gg 1$, one has
from (\ref{eq:2.9}) that $D^{E} \gg D$. On the other hand, this 
result do not mean that molecular diffusion $D$ plays no role in the
dispersion process. Indeed, if $D=0$ there is not mechanism for the
particles to exit from vortices.

Diffusion equation (\ref{eq:2.7}) 
is the typical long-time behavior in generic flow.
There exist also the possibility of the so-called
anomalous diffusion, i.e. when the spreading of
particle do not grow linearly in time, but with
a power law
\begin{equation}
\langle x^2(t) \rangle \sim t^{2\nu}
\label{eq:2.10}
\end{equation}
with $\nu \neq 1/2$. The 
case $\nu>1/2$ (formally $D^E = \infty$)  
is called super-diffusion; sub-diffusion, i.e.  $\nu<1/2$ (formally $D^{E}=0$), 
is possible only for compressible velocity fields.

Super-diffusion arises when the Taylor argument for deriving 
(\ref{eq:2.6}) fails and formally $D^{E} \to \infty$. This can
be due to one of the following  
 mechanisms: 

\noindent
a) the divergence of $\langle {\bf v}^2_{L} \rangle$
(which is the case of {\it L\'evy flights}), or 

\noindent
b) the lack of
decorrelation and thus $T_L \to \infty$ ({\it L\'evy walks}).
The second case is more physical and it is 
related to the existence of strong correlations in the dynamics,
even at large times and scales.

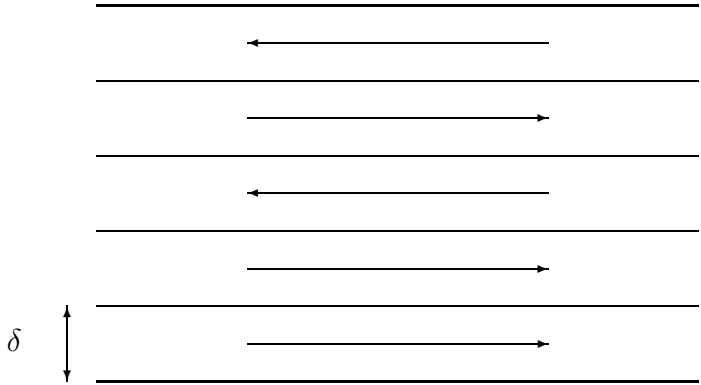
\begin{figure}[hbt]
\centerline{
\setlength{\unitlength}{2mm}
%\begin{picture}(40,10)(30,15)
\begin{picture}(40,10)(0,17)
\put(0,0){\line(1,0){40}}
\put(0,5){\line(1,0){40}}
\put(0,10){\line(1,0){40}}
\put(0,15){\line(1,0){40}}
\put(0,20){\line(1,0){40}}
\put(0,25){\line(1,0){40}}
\put(-2,0){\vector(0,1){5}}
\put(-2,5){\vector(0,-1){5}}
\put(-6,2){$\delta$}
\put(10,2.5){\vector(1,0){20}}
\put(10,7.5){\vector(1,0){20}}
\put(30,12.5){\vector(-1,0){20}}
\put(10,17.5){\vector(1,0){20}}
\put(30,22.5){\vector(-1,0){20}}
\end{picture}
}
\vspace{3.2cm}
\caption{Random shear of $\pm v_0$ velocity in strips of size $\delta$}
\label{fig4}
\end{figure}

One of the simplest examples of L\'evy walks is the dispersion
in a quenched random shear flow \cite{BG90,I92}. The flow,
sketched in Figure~\ref{fig4}, is a super-position of strips of
size $\delta$ of constant velocity $v_0$ with random directions.

Let us now consider a particle which moves according to the flow 
 of Figure \ref{fig4}.
Because the velocity field is in the $x$ direction only, in a time $t$ 
the typical displacement in the $y$ direction is due to molecular
diffusion only
\begin{equation}
\delta y \sim \sqrt{D t}
\label{eq:2.11}
\end{equation}
and thus in this time the walker visits $N=\delta y/\delta$ strips. 
Because of the
random distribution of the velocity in the strips, the mean velocity
in the $N$ strips is zero, but we may expect about $\sqrt{N}$ 
unbalanced strips (say in the right direction). 
The fraction of time $t$ spent in the unbalanced
strips is $t \sqrt{N}/N$ and thus we expect a displacement 
\begin{equation}
\delta x \sim v_0 {t \over \sqrt{N}} \, .
\label{eq:2.12}
\end{equation}
From (\ref{eq:2.11}) we have $N \sim \sqrt{D t}/\delta$ and finally
\begin{equation}
\langle \delta x^2 \rangle \sim {v_0^2 \delta \over \sqrt{D}} t^{3/2}
\label{eq:2.13}
\end{equation}
i.e. a super-diffusive behavior with exponent $\nu=3/4$.

The origin of the anomalous behavior in the above example is in the
quenched nature of the shear and in the presence of large stripes 
with positive (or negative) velocity in the $x$ direction. 
This leads to an infinite decorrelation
time for Lagrangian tracers and thus to a singularity in
(\ref{eq:2.6}). We conclude this example by observing 
that for $D \to 0$ (\ref{eq:2.13}) gives 
$\langle \delta x^2 \rangle \to \infty$. This is not a surprise because
in this case the motion is ballistic and the correct exponent
becomes $\nu=1$.

\begin{figure}[hbt]
\epsfxsize=12truecm
\centerline{\epsfbox{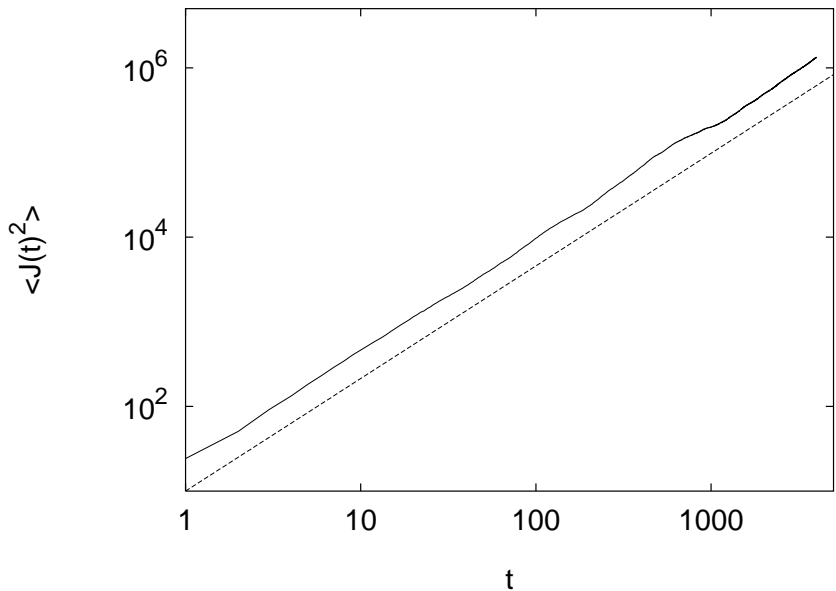}}
\caption{Square dispersion $\langle J(t)^2 \rangle$ for 
the standard map at $K=6.9155$. The dashed line is $t^{1.33}$.}
\label{fig5}
\end{figure}

As it was in the case of standard diffusion, also in the case of
anomalous diffusion the key ingredient is not randomness. Again, the
standard map model (\ref{eq:4.16}) is known to show anomalous behavior
for particular values of $K$ \cite{V98}. 
An example is plotted in Figure~\ref{fig5}
for $K=6.9115$ in which one find $\langle J(t)^2 \rangle \sim t^{1.33}$.

The qualitative mechanism for anomalous dispersion in the standard map can be
easily understood: a trajectory of (\ref{eq:4.16}) for which 
$K \sin \theta^{*} = 2 \pi m$
with $m$ integer, corresponds to a fixed point for $\theta$ (because the
angle is defined modulo $2 \pi$) and linear growth for $J(t)$ (ballistic
behavior). It can be shown that the stability region of these trajectories 
in phase space decreases as $1/K$ \cite{V98,IKH87} and, for intermediate 
value of $K$, they play a important role in transport: particles close
to these trajectories feel very long correlation times and perform very
long jumps. The contribution of these trajectory, as a whole, gives 
the observed anomalous behavior.

Now, let us consider the cellular flow of Figure~\ref{fig3} as
an example of sub-diffusive transport. 
We have seen that asymptotically (i.e. for $t \gg l_0^2/D$)
the transport is diffusive with effective diffusion coefficient 
which scales according to (\ref{eq:2.9}).
For intermediate times $l_0/v_0 \ll t \ll l_0^2/D$, when the 
boundary layer structure has set in, one expects 
anomalous sub-diffusive behavior as a consequence of fraction of
particles which are trapped inside vortices \cite{YPP89}.
A simple model for this problem is the comb model \cite{I92,HB87}:
a random walk on a lattice with comb-like geometry. 
The base of the comb represents the boundary layer of size 
$\delta$ around vortices
and the teeth, of length $l_0$, represent the inner area of the
convective cells. For the analogy with the flow of Figure~\ref{fig3}
the teeth are placed at the distance 
$\delta \sim \sqrt{D l_0/v_0}$ (\ref{eq:2.8}).

\begin{figure}[hbt]
\centerline{
\setlength{\unitlength}{2mm}
\begin{picture}(30,10)(0,4)
\put(0,0){\line(1,0){30}}
\put(0,0){\line(0,1){10}}
\put(3,0){\line(0,1){10}}
\put(6,0){\line(0,1){10}}
\put(9,0){\line(0,1){10}}
\put(12,0){\line(0,1){10}}
\put(15,0){\line(0,1){10}}
\put(18,0){\line(0,1){10}}
\put(21,0){\line(0,1){10}}
\put(24,0){\line(0,1){10}}
\put(27,0){\line(0,1){10}}
\put(30,0){\line(0,1){10}}
\put(1.5,-2){$\delta$}
\put(-2,5){$l_0$}
\end{picture}
}
\vspace{1.0cm}
\caption{The comb model geometry}
\label{fig6}
\end{figure}
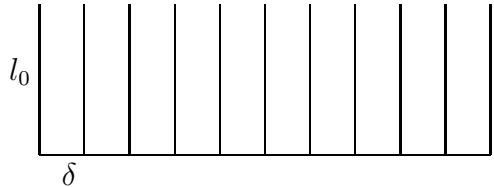

A spot of random walker (dye) is placed, at time $t=0$, at the base 
of the comb.
In their walk on the $x$ direction, the walkers can be trapped into
the teeth (vortices) of dimension $l_0$. 
For times $l_0/v_0 \ll t \ll l_0^2/D$, the dye invades a distance
of order $(D t)^{1/2}$ along the teeth. The fraction $F(t)$ of active dye
on the base (i.e. on the separatrix) decreases with time as
\begin{equation}
F(t) \sim {\delta \over (D t)^{1/2}}
\label{eq:2.14}
\end{equation}
and thus the effective dispersion along the base coordinate $b$ is
\begin{equation}
\langle b^2(t) \rangle \sim F(t) D t \sim \delta (D t)^{1/2}
\label{eq:2.15}
\end{equation}
In the physical space the corresponding displacement will be
\begin{equation}
\langle x^2(t) \rangle \sim \langle b^2(t) \rangle {l_0 \over \delta}
\sim l_0 (Pe D t)^{1/2}
\label{eq:2.16}
\end{equation}
i.e. we obtain a sub-diffusive behavior with $\nu=1/4$.

The above argument is correct only for the case of free-slip boundary
conditions. In the physical case of no-slip boundaries, one
obtains a different exponent $\nu=1/3$ \cite{YPP89}. 
The latter behavior has been indeed observed
in experiments \cite{CT88}.

%%%%%%%%%%%%%%%%%%%%%%%%%%%%%%%%%%%%%%%%%%%%%%%%%%%%%%%%%%%%%%%%%%%
%%%%%%%%%%%%%%%%%%%%%%%%%%%%%%%%%%%%%%%%%%%%%%%%%%%%%%%%%%%%%%%%%%%
%%%%%%%%%%%%%%%%%%%%%%%%%%%%%%%%%%%%%%%%%%%%%%%%%%%%%%%%%%%%%%%%%%%
\subsection{Beyond the diffusion coefficient}

From the above discussion it is now evident that diffusion,
being an asymptotic behavior, needs large scale separation in
order to be observed. In other words, diffusion arises only if
the Lagrangian correlation time $\tau_{L}$ (\ref{eq:2.5}) is finite 
{\it and} the observation time is $t \gg \tau_{L}$ or, according 
to (\ref{eq:2.7}), if the dispersion is evaluated on scales much larger than
$l_0$.

On the other hand, there are many physical and engineering applications
in which such a scale separation is not achievable. A typical 
example is the presence of boundaries which limit the scale of motion
on scales $L \sim l_0$. 
In these cases, it is necessary to introduce non-asymptotic quantities 
in order to correctly describe dispersion.

\begin{figure}[hbt]
\epsfxsize=12truecm
\centerline{\epsfbox{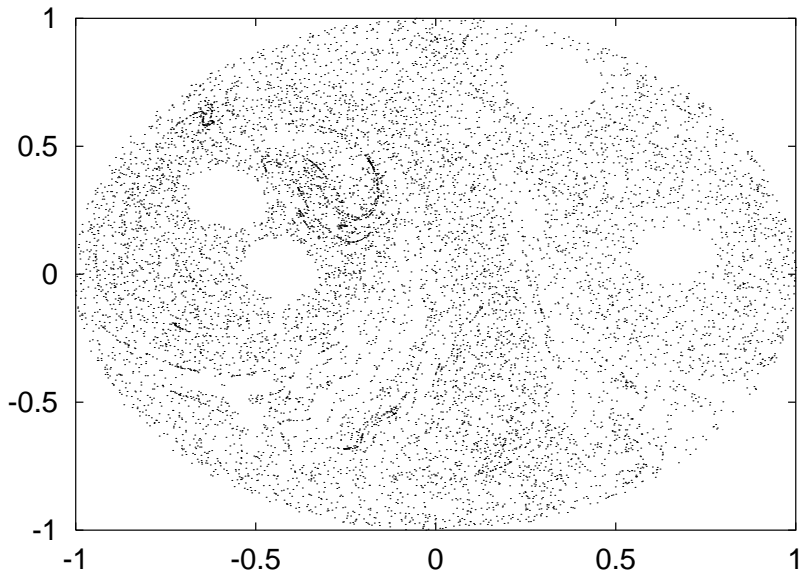}}
\caption{Position of the $10000$ tracers at a late stage
of the point vortex advection dynamics. The position of 
the 4 vortices are in the center of the ``hole''
in which the tracers cannot enter.}
\label{fig7}
\end{figure}

Before discussing the non-asymptotic statistics let us show, with an
example, how it can be dangerous to apply the standard analysis
in non-asymptotic situation. We consider the motion of tracers advected
by the two-dimensional flow generated by $4$ point vortices in a 
disk. The evolution equation is given by (\ref{eq:49}) and (\ref{eq:51})
but now in (\ref{eq:51}), instead of $\ln r_{ij}$, one has to consider 
the Green function $G(r_{ij})$ on the disk \cite{Lin41}.

A set of $10000$ tracers are initially placed in a very small cloud in 
the center of the disk. Because of the chaotic advection induced
by the vortices, at large time we observe the tracers dispersed in
all the disk (Figure~\ref{fig7}).

In the following we will consider {\it relative} dispersion,
i.e. the mean size of a cluster of particles
\begin{equation}
R^2(t) = \langle |{\bf x}(t) - 
\langle {\bf x}(t) \rangle|^2 \rangle 
\label{eq:3.1}
\end{equation}
Of course, for separation larger than the typical scale of the 
flow, $l_0$, the particles move independently and thus we 
expect again the asymptotic behavior
\begin{equation}
R^2(t) \simeq 2 D t \hspace{1cm} \mbox{if} \hspace{0.5cm} 
R^2(t)^{1/2} \gg l_0
\label{eq:3.2}
\end{equation}

For very small separation we expect, 
assuming that the Lagrangian motion is chaotic,
\begin{equation}
R^2(t) \simeq R^2(0) e^{2 \lambda t} \hspace{1cm} \mbox{if} 
\hspace{0.5cm} R^2(t)^{1/2} \ll l_0
\label{eq:3.3}
\end{equation}
where $\lambda$ is the Lagrangian Lyapunov exponent \cite{CFPV91}.

\begin{figure}[hbt]
\epsfxsize=12truecm
\centerline{\epsfbox{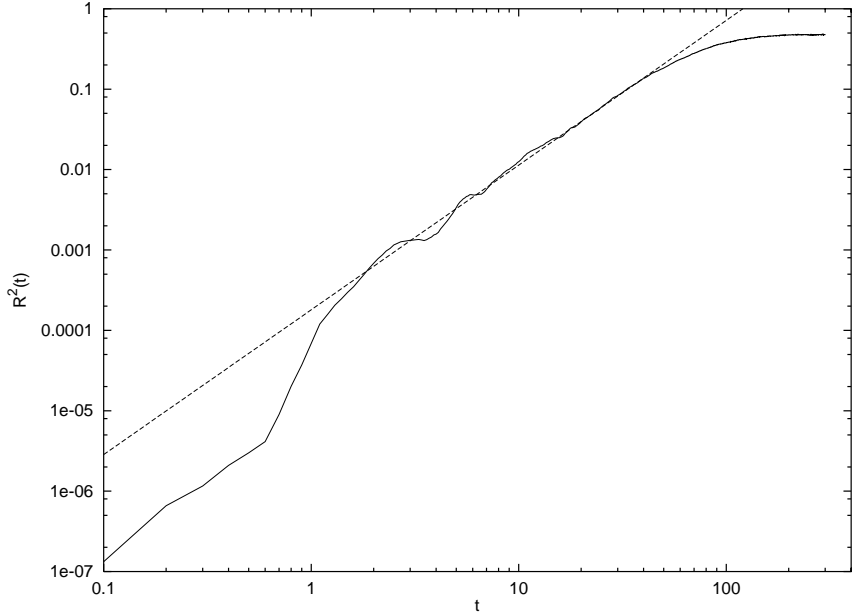}}
\caption{$R^2(t)$ computed for the four 
point vortex system. The dashed line is the power law
$\langle R^2(t) \rangle \sim t^{1.8}$.}
\label{fig8}
\end{figure}

The computation of the standard dispersion for the
tracers in the point vortex model is plotted in Figure~\ref{fig8}. 
At very long time $R^2(t)$ reaches the saturation value
due to the boundary.

For intermediate times a power-law behavior 
with an anomalous exponent $\nu=1.8$ is clearly observable. 
Of course the anomalous behavior is spurious: after the 
discussion of the previous section, we do not see any reason
for observing super-diffusion in the point vortex system.
The apparent anomaly is simply due to the lack of scale 
separation and thus to the crossover from the exponential
regime (\ref{eq:3.3}) to the saturation value.

To partially avoid this kind of problem, it has been recently
introduced a new indicator based on {\it fixed scale} analysis 
\cite{ABCCV97}.
The idea is very simple and it is based on exit time statistics.
Given a set of thresholds $\delta_n=\delta_0 r^n$, one measures
the exit time $T_{i}(\delta_n)$ it takes for the separation $R_{i}(t)$
to grow from $\delta_n$ to $\delta_{n+1}$. The factor $r$ may
be any value $>1$, but it should be not too large in order to have 
a good separation between the scales of motion. 

Performing the exit time experiment over $N$ particle pairs, from
the average doubling time 
$\langle T(\delta) \rangle = 1/N \sum_{i} T_i(\delta)$, one defines
the Finite Size Lyapunov Exponent (FSLE) as
\begin{equation}
\lambda(\delta) = {\ln r \over \langle T(\delta) \rangle}
\label{eq:3.4}
\end{equation}
which recovers the standard Lagrangian Lyapunov exponent in the 
limit of very small separations
$\lambda = \lim_{\delta \to 0} \lambda(\delta)$.

The finite size diffusion coefficient $D(\delta)$ is defined,
within this framework, as
\begin{equation}
D(\delta) = \delta^2 \lambda(\delta)
\label{eq:3.5}
\end{equation}
For standard diffusion $D(\delta)$ approaches the 
diffusion coefficient $D$ (see (\ref{eq:3.2})) in the limit
of very large separations ($\delta \gg l_0$). This result stems from
the scaling of the doubling times 
$\langle T(\delta) \rangle \sim \delta^2$ 
for normal diffusion.

Thus, according to (\ref{eq:3.2})-(\ref{eq:3.3}), the asymptotic
behaviors of the FSLE are
\begin{equation}
\lambda(\delta) \sim \left\{
\begin{array}{ll}
\lambda & \hspace{1cm} \mbox{if} \hspace{0.5cm} \delta \ll l_0 \\
D/\delta^{2} & \hspace{1cm} \mbox{if} \hspace{0.5cm} \delta \gg l_0
\end{array}
\right.
\label{eq:3.6}
\end{equation} 

In presence of boundary at scales $L \sim l_0$, the second regime 
is not observable. For separation very close to to the saturation
value $\delta_{max} \simeq L$ one expects the following behavior to
hold for a broad class of systems \cite{ABCCV97}:
\begin{equation}
\lambda(\delta) \propto {\delta_{max} - \delta \over \delta}
\label{eq:3.7}
\end{equation}

\begin{figure}[hbt]
\epsfxsize=12truecm
\centerline{\epsfbox{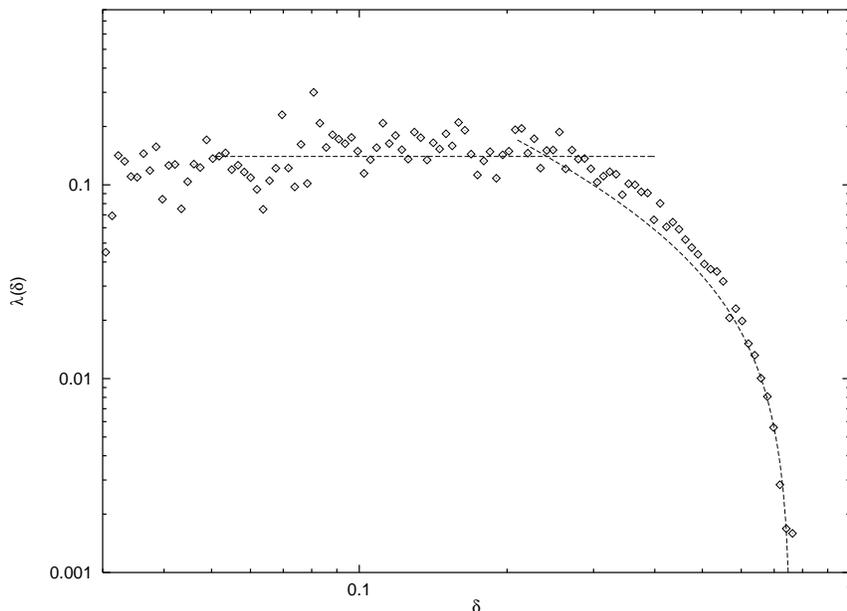}}
\caption{FSLE $\lambda(\delta)$ for the tracers advected by the 
point vortex system. The horizontal line represent the Lagrangian
Lyapunov exponent ($\lambda \simeq 0.14$). The dashed curve is
the saturation regime with $\delta_{max}=0.76$.}
\label{fig9}
\end{figure}

Let us now come back to the point vortex example of Figure~\ref{fig7}.
The FSLE for this problem is plotted in Figure~\ref{fig9}.

With the finite scale analysis one clearly see that only two
regime survive: exponential at small scales (chaotic advection)
and saturation at large scale. The apparent anomalous regime 
of Figure~\ref{fig8} is a spurious effect induced by taking the
average at fixed time.

The finite scale method can be easily applied to the analysis
of experimental data \cite{BCEQ00}. 
An example is the study of Lagrangian
dispersion in a experimental convective cell. The cell is a
rectangular tank filled with water and heated by a linear
heat source placed on the bottom. The heater generates a 
vertical plume which induces a general two-dimensional circulation 
of two counter-rotating vortices. For high values of the Rayleigh
number (i.e. heater temperature) the flow is not stationary and
the plume oscillates periodically. 
In these conditions, Lagrangian tracers can jump from one side 
to the other of the plume as a consequence of chaotic advection.

\begin{figure}[hbt]
\epsfxsize=12truecm
\centerline{\epsfbox{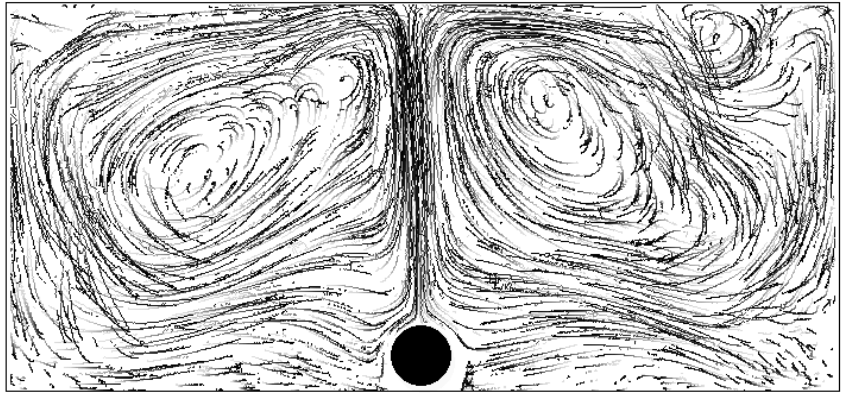}}
\caption{An example of trajectories obtained by PTV technique 
in the convective cell at $Ra=2.39 \times 10^{8}$. The vertical
thermal plume is clearly observable. The dark circle on the bottom
represents the heat source.}
\label{fig10}
\end{figure}

\begin{figure}[hbt]
\epsfxsize=12truecm
\centerline{\epsfbox{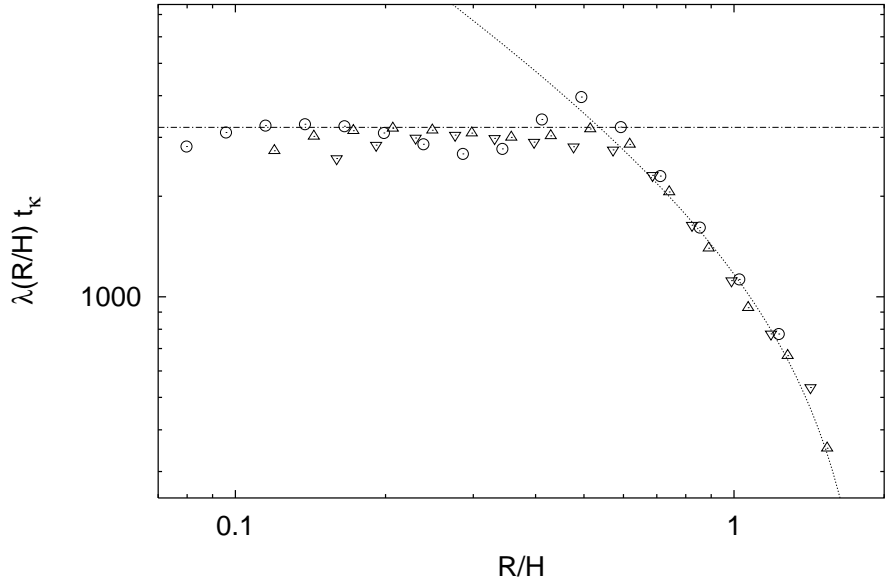}}
\caption{FSLE $\lambda(\delta)$ computed for the convective cell
at different initial separations (different symbols). The straight 
line is the Lyapunov exponent and the dashed curve represents 
the saturation regime.}
\label{fig11}
\end{figure}

The study of Lagrangian dispersion has been done by means
of the FSLE \cite{BCEQ00}. In Figure~\ref{fig11} we plot
the result for $Ra=2.39 \times 10^{8}$. Again, because there
is no scale separation between the Eulerian characteristic 
scale $l_0$ (vortex size) and the basin scale $L$ we cannot
expect diffusion behavior. Indeed, the FSLE analysis reveals 
the chaotic regime $\lambda(\delta)=\lambda$ at small scales
and the saturation regime (\ref{eq:3.7}) at larger scale.

The finite scale tool has been successfully applied to many other
numerical and experimental situations, from the dispersion
in fully developed turbulence, to the analysis of tracer motion in 
 ocean and atmosphere \cite{LAV01,boffi01,legras02}, 
to engineering laboratory experiments.
It will be probably became a standard tool in the analysis of 
Lagrangian dispersion.

%%%%%%%%%%%%%%%%%%%%%%%%%%%%%%%%%%%%%%%%%%%%%%%%%%%%%%%%%%%%%%%%%%%

\end{document}